\documentclass[a4paper,11pt]{article}
\pdfoutput=1 %

\usepackage{jcappub} %

\usepackage{aas_macros}    %
\usepackage[utf8]{inputenc}
\usepackage{textgreek}
\usepackage[T1]{fontenc} %
\usepackage[dvipsnames]{xcolor}
\usepackage[normalem]{ulem}

\newcommand{\lya}{Ly$\alpha$}

\newcommand{\invMpc}{\mathrm{Mpc}^{-1}}

\newcommand{\hinvMpc}{h \invMpc}

\renewcommand{\vec}{\mathbf}

\newcommand{\HeII}{\textsc{He II}}

\title{\boldmath Simulating intergalactic gas for DESI-like small scale Lyman$\alpha$ forest observations}

\author[a,1]{Michael Walther\note{Corresponding author},}
\author[a]{Eric Armengaud,}  
\author[a]{Corentin Ravoux,}
\author[a]{Nathalie Palanque-Delabrouille,}
\author[a]{Christophe Y\`eche,}
\author[b]{Zarija Luki\'c}

\affiliation[a]{IRFU, CEA, Universit\'e Paris-Saclay, F-91191 Gif-sur-Yvette}
\affiliation[b]{Lawrence Berkeley National Laboratory, Berkeley, California}

\emailAdd{michael.walther@cea.fr}
\emailAdd{eric.armengaud@cea.fr}
\emailAdd{corentin.ravoux@cea.fr}
\emailAdd{nathalie.palanque-delabrouille@cea.fr}
\emailAdd{christophe.yeche@cea.fr}
\emailAdd{zarija@lbl.gov}

\arxivnumber{2012.04008}

\abstract{
Measurements of the \lya{} forest based on large numbers of quasar spectra from sky surveys such as SDSS/eBOSS accurately probe the distribution of matter on small scales and thus provide important constraints on several ingredients of the cosmological model. A main summary statistic derived from those measurements is the one-dimensional power spectrum, $P_{\rm 1D}$, of the \lya{} absorption. However, model predictions for $P_{\rm 1D}$ rely on expensive hydrodynamical simulations of the intergalactic medium, which was the limiting factor in previous analyses. Datasets from upcoming surveys such as DESI will push observational accuracy near the 1$\%$-level and probe even smaller scales. This observational push mandates even more accurate simulations as well as more careful exploration of parameter space. In this work we evaluate the robustness and accuracy of simulations and the statistical framework used to constrain cosmological parameters. We present a comparison between the grid-based simulation code Nyx and SPH-based code Gadget in the context of $P_{\rm 1D}$. In addition, we perform resolution and box-size convergence tests using Nyx code.
We use a Gaussian process emulation scheme to reduce the number of simulations required for exploration of parameter space without sacrificing the model accuracy. We demonstrate the ability to produce unbiased parameter constraints in an end-to-end inference test using mock eBOSS- and DESI-like data, and we advocate for the usage of adaptive sampling schemes as opposed to using a fixed Latin hypercube design.
}

\begin{document}
\maketitle
\flushbottom

\section{Introduction}
\label{sec:intro}

Observations of the large-scale structure of the Universe are essential for addressing a number of outstanding questions in cosmology. At redshifts $2<z<6$ the \lya{} forest provides the primary probe of this structure on scales ranging from sub-Mpc to tens of Mpc. The \lya{} forest, a series of absorption features in the spectrum of bright background sources, is imprinted on the spectra of distant background sources by absorption from neutral hydrogen located in the intergalactic medium (IGM). It complements information from other probes, such as the cosmic microwave background (CMB) or galaxy clustering, especially by extending the reach to significantly smaller scales.  This allows extracting information on the physics of the IGM, neutrino masses, dark matter as well as inflation.

The exploration of neutrino properties is a major window in the search for physics beyond the standard model of particle physics.
Neutrino oscillation experiments~\cite{deSalas2018} have indicated that neutrinos are massive, and provide a lower limit to the sum of neutrino masses: $\sum m_i > 0.06\,\mathrm{eV}$.
However, despite intense effort from particle physics (kinematic $\beta$ decay experiments, e.g.~\cite{Aker:2020}, neutrinoless double-$\beta$ decay experiments, e.g.~\cite{Agostini:2019}) and cosmology (e.g.~\cite{planckcollaborationPlanck2018Results2018a}), the absolute neutrino mass scale remains unknown making its measurement a major challenge today. 
From an astrophysical point of view, neutrinos are among the most abundant particles in the Universe.
Thus, even a 0.06 eV total neutrino mass has a notable impact on several probes of cosmological structures, from CMB anisotropies to the clustering of galaxies and the distribution of hydrogen in the IGM.
One of the most sensitive approaches to measuring neutrino mass lies in combining observations of the CMB, which probe the largest observable structures of the universe, with measurements of the matter power spectrum on scales of a few Mpc, like the \lya{} forest. Indeed, while large scales are mostly unaffected by neutrino masses, Mpc-scale fluctuations are partly suppressed due to massive neutrinos. This approach has already provided increasingly strong constraints on the sum of neutrino masses~\cite{Croft:1999, Seljak:2005, Palanque-Delabrouille:2015a, Palanque-Delabrouille:2015b, Yeche:2017, palanque-delabrouille2019}. 

The nature of dark matter is yet another long-standing mystery in cosmology.
While first hints for dark matter are almost a century old~\cite{Zwicky:1933}, there is still no indication what its nature might be.
Weakly interacting massive particles in the 100 GeV mass range have long been considered as the most viable candidates, but direct detection or searches led at accelerators have excluded a large number of plausible models~\cite{Aprile:2017}.
Other candidates are now getting more and more attention from the scientific community, and their impact needs to be studied on all fronts.
In particular, some models do predict that cosmological structure is smoothed out on scales of a few Mpc. This is the case of ultra-light bosons with a mass $\sim 10^{-22}$~eV~\cite{Hui:2017}, and sterile neutrinos in the few keV range. The later have also been claimed as an explanation for observations of a $3.5\,\mathrm{keV}$ emission line from dark matter-dominated objects~\cite{Boyarski:2014}. On the contrary, other dark matter models predict an enhancement of Mpc-scale structures. All these scenarios are best probed by \lya{} data~\cite{Narayanan:2000, Seljak:2006, Viel:2013, Baur:2016, Baur:2017, Armengaud:2017, Irsic:2017, Murgia:2018, palanque-delabrouille2019} (for a recent analysis using other tracers, based on e.g. gravitational lensing, see~\cite{Enzi:2020}).

Last but not least, the cosmological concordance model assumes an inflationary phase early on in the history of the universe.
While successful in providing an elegant solution to fundamental problems of cosmology, the detailed physics of such a phase remains largely elusive.
Therefore, observations are required to characterize the underlying physical processes behind inflation. 
Inflation predicts that the power spectrum of matter fluctuations in the primordial Universe is a power-law to first order, and the exact value of the power law index $n_s$ as well as possible small departures (the so-called running of this index) depend on the inflationary scenario.
While CMB anisotropies already provide a precise measurement of $n_s$ on large scales, probing the matter power spectrum at Mpc to tens of Mpc scales grants a very powerful lever arm to constrain its scale dependence~\cite{Phillips:2001,Hannestad:2002,Viel:2004,Seljak:2005,Palanque-Delabrouille:2015b}.

However, inferring physical and cosmological parameters from small-scale properties of the  \lya{} forest is not straightforward.
For redshifts ($z\sim 2-4$) and scales ($1 - 10$~Mpc) where most measurements are done, the gravitational growth of structures is in a mildly non-linear regime.
In addition, the small-scale structure of the \lya{} forest is strongly affected by intrinsic properties of the gas, e.g. its temperature and pressure support, which itself depend on the evolution of the underlying UV radiation field \cite{Hui:1997}. 
The \lya{} forest thus cannot be considered a simple biased tracer of the dark matter on those scales.
Cosmological interpretations of small-scale \lya{} forest data require hydrodynamical simulations to accurately model the evolution of IGM properties as the cosmic structures evolve with redshift. 

Currently, the largest amount of \lya{} data come from the SDSS/BOSS/eBOSS surveys~\cite{McDonald:2006,Palanque-Delabrouille:2013,Chabanier:2019}.
They already provide a precise measurement of the one-dimensional \lya{} flux absorption power spectrum $P_{\rm 1D}(k)$:~\cite{Chabanier:2019} uses spectra from 180,413 quasars to measure $P_{\rm 1D}(k)$ to $\sim 1\text{--}5\%$ accuracy over a range of scales $10^{-3} \lesssim k \lesssim 2\times 10^{-2}\,\mathrm{s/km}$ (corresponding to $0.075\lesssim k \lesssim 1.5\,\mathrm{Mpc^{-1}}$), and redshifts $z = 2.1-4.7$.
SDSS data are complemented by observations at higher resolution constraining even smaller scales (larger $k$) and  different regimes in redshift~\cite{McDonald:2000,Croft:2002,Kim:2004,Viel:2013,Irsic:2017,Yeche:2017,Walther:2018,Khaire:2019, Boera:2019,Day:2019}.
While the high-resolution measurements improve the constraints on the thermal state of the IGM~\cite{Walther:2019, Boera:2019} (and can also be used for stronger constraints on e.g. warm dark matter (WDM) mass and reionization~\cite{Viel:2013,Irsic:2017}, but see~\cite{Garzilli:2019}), a large fraction of the cosmological information is encoded in SDSS data.
The cosmological interpretation of existing data was done in particular from a grid of hydrodynamical simulations, using the Gadget-3 software~\cite{Borde:2013,Palanque-Delabrouille:2015a,Palanque-Delabrouille:2015b,palanque-delabrouille2019}.
Those cosmological simulations are very demanding in terms of CPU usage. To overcome computing limitations, the so-called splicing technique was used which combines simulations over different scales to cover the range of scales $k$ of interest~\cite{Borde:2013}. Note that while probably good enough for previous analyses, the accuracy estimates of this approach have been doubted in other works, e.g.~\cite{Lukic:2015} and also led to the requirement of additional nuisance parameters which one marginalized over in the analyses~\cite{Palanque-Delabrouille:2015a,palanque-delabrouille2019}.

The recently commenced Dark Energy Spectroscopic Instrument (DESI)~\cite{DESIFinalDesignReport} and the upcoming WEAVE-QSO survey \cite{Pieri:2016} are expected to provide, among others, improved measurements of $P_{\rm 1D}(k)$. Thanks to both increased statistics (about 700,000 \lya{} forest quasar spectra) and improved instrumental performance (higher resolution, in particular) they should reach a near-\% level of precision and accuracy over an increased range of scales compared to previous surveys. 
A new generation of large cosmological simulations with comparable accuracy is therefore necessary to interpret this measurement. 

In this work we perform simulations with the cosmological hydrodynamical grid code Nyx~\cite{Almgren:2013,Lukic:2015}. A major advantage of this software is the reduction of computational costs allowing to run simulations with a higher dynamic range --- and thus ideally higher accuracy --- with similar computational allocations.
We compare predictions from a fiducial Nyx simulation to those obtained from the Gadget-3 code, which uses Smoothed Particle Hydrodynamics (SPH) to model the IGM, assessing the differences between both approaches in predicting the \lya{} forest. Note that for similar simulation setups, the Nyx runs were completed in about $10\%$ of the time required for the Gadget run, and that Nyx is currently ported to GPUs allowing large additional boosts of computational speed in the near future. 
While similar comparisons were done before (e.g.~\cite{regan2007} comparing Gadget and Enzo), those were performed on smaller simulations that were less resolved and thus did not compare simulations to the required accuracy.
Our main focus is to determine the impact of the choice of simulation code on the flux power spectrum. 

Then, we study the convergence of Nyx simulations as a function of simulation volume and resolution to assess the minimal requirements allowing successful analyses on DESI data (based on the results of \cite{Lukic:2015}).
Finally, in order to optimally interpolate between predictions from simulations with different parameters, we present a test for novel emulation schemes based on Gaussian processes (see~\cite{birdEmulatorLymanalphaForest2018} for a similar approach).
The purpose of this study lies in defining the requirements for a large grid of cosmological simulations, to interpret DESI \lya{} forest data, and to replace previously existing grids that have been used to build Taylor expansions of the power spectrum for BOSS and eBOSS analyses~\cite{Borde:2013,Rossi:2014}. For a final analysis those new grids are envisioned to be supplemented by refinement simulations in a Bayesian optimization approach as described e.g. in~\cite{rogersBayesianEmulatorOptimisation2018,takhtaganov:2019}.

This paper is structured in the following way. In \autoref{sec:sims} we discuss the different simulations used in this analysis and present a comparison of their flux power spectra. The requirements for the simulations with respect to the box size and resolution are presented in \autoref{sec:requirements}. In \autoref{sec:emulator} we present our approach to cosmic emulation, demonstrating its robustness and accuracy.  We conclude our work in \autoref{sec:conclusion}.

\section{Simulation approach}\label{sec:sims}

In this section we introduce the simulation codes used in our comparison between SPH (using Gadget~\cite{Springel:2005} and MP-Gadget~\cite{mpgadget-cite}) and mesh based (using Nyx~\cite{Almgren:2013, Lukic:2015}) hydrodynamics. We also introduce the fiducial cosmological and astrophysical parameters used.

\subsection{Cosmological scenario and initial conditions}

\begin{table}[!ht]
    \caption{Input cosmological parameters for the code comparison}
    \label{tab:cc_params}
    \centering
    \begin{tabular}{cc}
        \hline\hline
        $\Omega_\mathrm{m}$ & 0.3144 \\
        $\Omega_\mathrm{b}$ & 0.04938 \\
        $n_s$ & 0.9660 \\
        $h$ & 0.6732 \\
        $m_\nu$ & 0 \\
        $A_s$ & $2.101\cdot 10^{-9}$\\
        $\sigma_8$ (derived)& 0.8235 \\
        \hline\hline
    \end{tabular}\\
\end{table}

We define a benchmark scenario based on the Planck 2018~\cite{planckcollaborationPlanck2018Results2018a} cosmological results, but without neutrinos, i.e. we simulate a flat, neutrinoless Universe with the parameters given in \autoref{tab:cc_params}.

The early-time statistical description of density fluctuations for the different fluids in the Universe is described by transfer functions. They are computed in the linear regime at the initial simulation redshift $z_\mathrm{ini} = 99$, making use of \texttt{CAMB}\footnote{\url{https://www.camb.info}}~\cite{Lewis:2000}.
Particle realizations are then generated at $z_\mathrm{ini}$ in second order Lagrangian perturbation theory (2LPT) using \texttt{2lpt-ic} \footnote{\url{https://cosmo.nyu.edu/roman/2LPT/}}~\cite{Crocce:2012}. 
For the SPH codes (Gadget-3, MP-Gadget) we generated 2 particle species following the dark matter and baryon transfer functions\footnote{note that the approach used for this task within \texttt{2lpt-ic} is not ideal and will lead to unphysical changes in the growth of Baryons and dark matter at later times. Possible solutions to this issue have been developed by~\cite{Bird:2020} and~\cite{Hahn:2020}, but have not yet been used for this work.} and used them as initial conditions (ICs) for both particle species. 
For Nyx the standard approach is to use a single transfer function for the total matter to generate dark matter particles. 
Nyx then generates the initial baryon grid by subtracting a fraction of their mass from the particles and depositing it on a mesh using a cloud-in-cell (CIC) scheme.
Most of our runs were performed using this approach.
As an alternative, Nyx also allows reading a second IC file with gas particles and depositing them to the grid via CIC. 
We ran one such simulation with the identical particle realizations as for the SPH codes to assess the impact of the different initializations on the flux power.
We choose common random seeds for both types of simulation in order to minimize the impact of {sample} variance on our comparison. 

In our benchmark scenario, we used a common set of assumptions to describe the physical evolution of the IGM in all simulations. A constant He fraction $Y=0.24$ is assumed, with no metals. The effect of a time-varying, but spatially uniform, ultraviolet background (UVB) on photoionization and photoheating is taken into account by giving to the codes a list of redshift-dependent rates~\cite{Katz:1992}. These rates were derived from the late reionization model of~\cite{Onorbe:2018}, and a warm \HeII{} reionization scenario with $\Delta T=20000K$, whose resulting thermal evolution is in good agreement with the measurement of~\cite{Walther:2019}.
Effects of inhomogeneous reionization are therefore neglected, both in terms of temperature and UV background fluctuations.
While we expect future simulations to increasingly use models with fluctuations \citep{Onorbe:2019} in order to achieve better accuracy of the power spectrum, here we want to compare different codes in their most commonly used regime. The thermal state of the IGM is first of all described by the temperature-density scaling relation $T = T_0 \, \Delta^{\gamma-1}$, measured for each simulation box. 
The thermal pressure smoothing of gas density fluctuations with respect to the dark matter distribution is commonly
parametrized by the pressure smoothing scale $\lambda_P$.
While long after any reionization event the instantaneous temperature is more or less independent of the redshift of reionization, $\lambda_P$ retains a memory of this for a longer time~\cite{Gnedin:2003,Kulkarni:2015, Onorbe:2018}.
Inputs for the codes are somewhat different (due to different initialization of baryons), and we  use different analysis toolchains adapted to each code. Our results therefore reflect the way an end-user would report results, and do not provide a comparison of the simulations on fully equal ground. {A list of the different simulations that have been run can be found in \autoref{tab:sim_properties}.}

\subsection{Used hydrodynamical codes}

\begin{table}[!ht]
    \caption{{Simulations 
    used in the paper. Most simulations have been run with the parameters of \autoref{tab:cc_params}. The exception are those dubbed "emulator" (see \autoref{sec:emulator} for details and \autoref{tab:lh_params} for their parameters). Outputs were produced in intervals of $\Delta z=0.2$ from $z=4.6$ until $z_f$, Note that in the core of the paper, boxsize values have been rounded to multiples of $10\,\mathrm{Mpc}$.  $N_{\mathrm gas}$ refers to the number of SPH particles for (MP)-Gadget, and is the number of cells in the mesh for Nyx.}}
    \label{tab:sim_properties}
    \centering
    \begin{tabular}{cccccc}
        \hline\hline
        &boxsize [Mpc]&$N_{gas}$&$N_{dm}$& transfer function &$z_f$\\
        \hline
        Gadget-III fiducial&29.71&$1024^3$&$1024^3$& gas and DM seperately&2.2\\
        MP-Gadget fiducial&29.71&$1024^3$&$1024^3$& gas and DM seperately&3.6\\
        Nyx fiducial&29.71&$1024^3$&$1024^3$& all matter&2.2\\
        Nyx (init with gas particles)&29.71&$1024^3$&$1024^3$& gas and DM seperately&2.2\\
        Nyx small&14.85&$512^3$&$512^3$& all matter&2.2\\
        Nyx small highres&14.85&$1024^3$&$1024^3$& all matter&2.6\\
        Nyx medium&59.42&$2048^3$&$2048^3$& all matter&2.2\\
        Nyx big &118.8&$4096^3$&$4096^3$& all matter&2.2\\
        Nyx emulator 15,512 &15&$512^3$&$512^3$& all matter&2.2\\
        Nyx emulator 60,512 &60&$512^3$&$512^3$& all matter&2.2\\
        Nyx emulator 15,128 &15&$128^3$&$128^3$& all matter&2.2\\
        \hline\hline
    \end{tabular}\\
\end{table}

\subsubsection{Gadget}
\label{sec:Gadget3}
Gadget-3 was the code most often used for cosmological interpretation of $P_{\rm 1D}(k)$ measurements so far \cite{Palanque-Delabrouille:2015a,palanque-delabrouille2019}. Note that also for astrophysical analysis of the IGM, reionization, and in the context of WDM constraints, many other groups use this code, e.g. in \cite{Becker:2011,Viel:2013,Bolton:2017,Boera:2019,Gaikwad:2020}. It is a massively parallel code for cosmological hydrodynamics and was last described in~\cite{Springel:2005}\footnote{Note that very recently a new version of the code was released in \cite{Springel:2020}, this however was too late to be included in our comparison.}. To follow the dynamics of both collisionless dark matter and cold, condensed baryonic matter (aka "stars") a N-body method is used. Hydrodynamics in the IGM is described via SPH. Gravitational interactions are computed with a TreePM method, so that the long-range part of this force is estimated with a particle-mesh algorithm. A cubic spline kernel is used for gravitational softening and the SPH density calculation, with a softening length adapted to the local particle density (1/40 mean particle separation).

As is common for \lya{} forest simulations, the physics of gas particles takes into account the main cooling and heating processes in which ionized and neutral Hydrogen and Helium are involved, as described, for example, in~\cite{Katz:1992}. In this work, we run Gadget with the QUICKLYA option.  With this option, gas particles are turned into "stars" using a highly simplified method, where all SPH particles above the overdensity of $\Delta=1000$ times the mean are converted into star particles.  We do not use any modeling of stellar or AGN feedback.

An important caveat of Gadget-3 based hydrodynamics simulations is its large time consumption, even when making use of the QUICKLYA option. Indeed, more than 100 kCPU hours are needed for a 20~Mpc-h$^{-3}$, $1024^3$ particle simulation to run until $z=2.2$ on 86 nodes of the Joliot Curie computer each containing 48 Intel Skylake cores. This is mostly driven by the late-time evolution, following the densest regions of gas, which are not the most relevant ones for the \lya{} forest (see \autoref{sec:density-dependence}).

In some cases, we also used an additional simulation based on the MP-Gadget code\footnote{available from \url{https://github.com/MP-Gadget/MP-Gadget/}, we used commit \texttt{7bade615}.}~\cite{mpgadget-cite,Feng:2014}, a heavily modified version of the P-Gadget3 code focusing especially on scalability and force accuracy.  
We used the cooling rates from~\cite{Katz:1992} as in the standard Gadget run, enabled the QUICKLYA procedure with the same threshold overdensity $\Delta=1000$ as for Gadget-3, disabled the stellar and AGN feedback models, and initialized with the same ICs (but converted to MP-Gadget format). {We only ran this simulation to $z=3.6$ due to limitations in computing time allocation. This was sufficient for the comparison purposes of this study.}
From Gadget or MP-Gadget outputs, matter power spectra and \lya{} flux-based statistics were computed using \texttt{GenPk}, \texttt{extract} and  \texttt{fake\textunderscore{}spectra}.

\subsubsection{Nyx}
\begin{figure}
        \centering
        \includegraphics[width=0.5\textwidth]{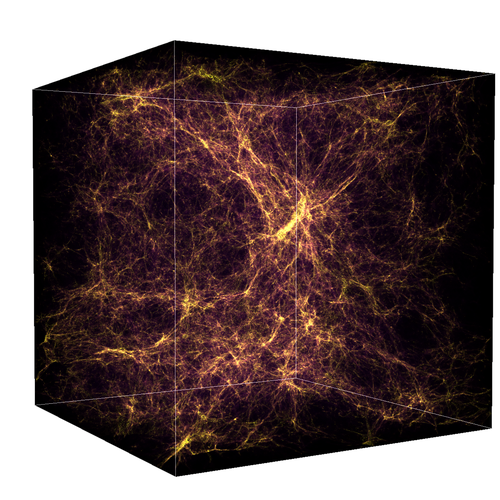}
        \caption{Volume rendering of the baryon density for the $2048^3$ Nyx box at $z=2.4$.}
        \label{fig:render}
\end{figure}

Nyx\footnote{available from \url{https://github.com/AMReX-Astro/Nyx}} follows the evolution of dark matter simulated as self-gravitating Lagrangian particles, and, in contrast to the previous codes, models baryons as an ideal gas on a uniform Cartesian grid.
Note that applying a mesh-based approach has some general advantages compared to using a SPH-code, such as a clearly defined spatial resolution (as opposed to the fixed mass resolution of SPH codes), and in general more accurate treatment of shocks and hydrodynamics.
An obvious disadvantage is that in the absence of adaptive mesh refinement (AMR) techniques, the dynamic range is limited by the grid size.
We do not use AMR here as the \lya{} forest gas spans most of the simulation box, making AMR techniques inefficient\footnote{Note that, e.g. AGN feedback also affects the IGM and requires accurate modeling of high-density regions as enabled by AMR (see e.g. \cite{Chabanier:2020} for a recent study), but these effects are beyond the scope of this work.}.
In Nyx, gas dynamics is solved using a second-order accurate finite volume methodology.
For more details of these numerical methods and scaling behavior tests, see~\cite{Almgren:2013} and~\cite{Lukic:2015}.
Besides solving for gravity and the Euler equations, we also include the main physical processes fundamental to model the Lyα forest.
First, we consider the chemistry of the gas as having a primordial composition with hydrogen and helium mass abundances of $X_p$ and $Y_p$ , respectively.
In addition, we include inverse Compton cooling off the microwave background and keep track of the net loss of thermal energy resulting from atomic collisional processes.
We use the updated recombination, collision ionization, dielectric recombination rates, and cooling rates given in~\cite{Lukic:2015}.
All cells are assumed to be optically thin to ionizing radiation, and radiative feedback is accounted for via the same uniform UVB as for the Gadget run. The results of this paper are based on Nyx version 18.05, with exception of the largest box in the boxsize convergence study where we used version 20.02.\footnote{{The main updates of the newer version are to allow new modes of parallelization and new features that have been implemented and which we do not use here. This study is therefore unaffected by the change.}}

All post-processing for Nyx was performed using the \texttt{gimlet} postprocessing suite (see, for example \cite{Friesen:2016}). 
This software package allows for efficient, MPI-parallel computation of typical statistical properties in a Nyx simulation based on the hydrodynamical grid and CIC depositions of the DM particles.

The properties of fiducial Nyx and Gadget simulation outputs in terms of thermal evolution and matter distribution are described in Appendices \ref{sec:sim-thermal-cmp} and \ref{sec:sim-matter-dist-cmp}.

\subsection{Comparison of the 1d Ly$\alpha$ flux power spectrum between simulations}\label{sec:results}

\begin{figure}
    \centering
    \includegraphics[width=0.9\textwidth]{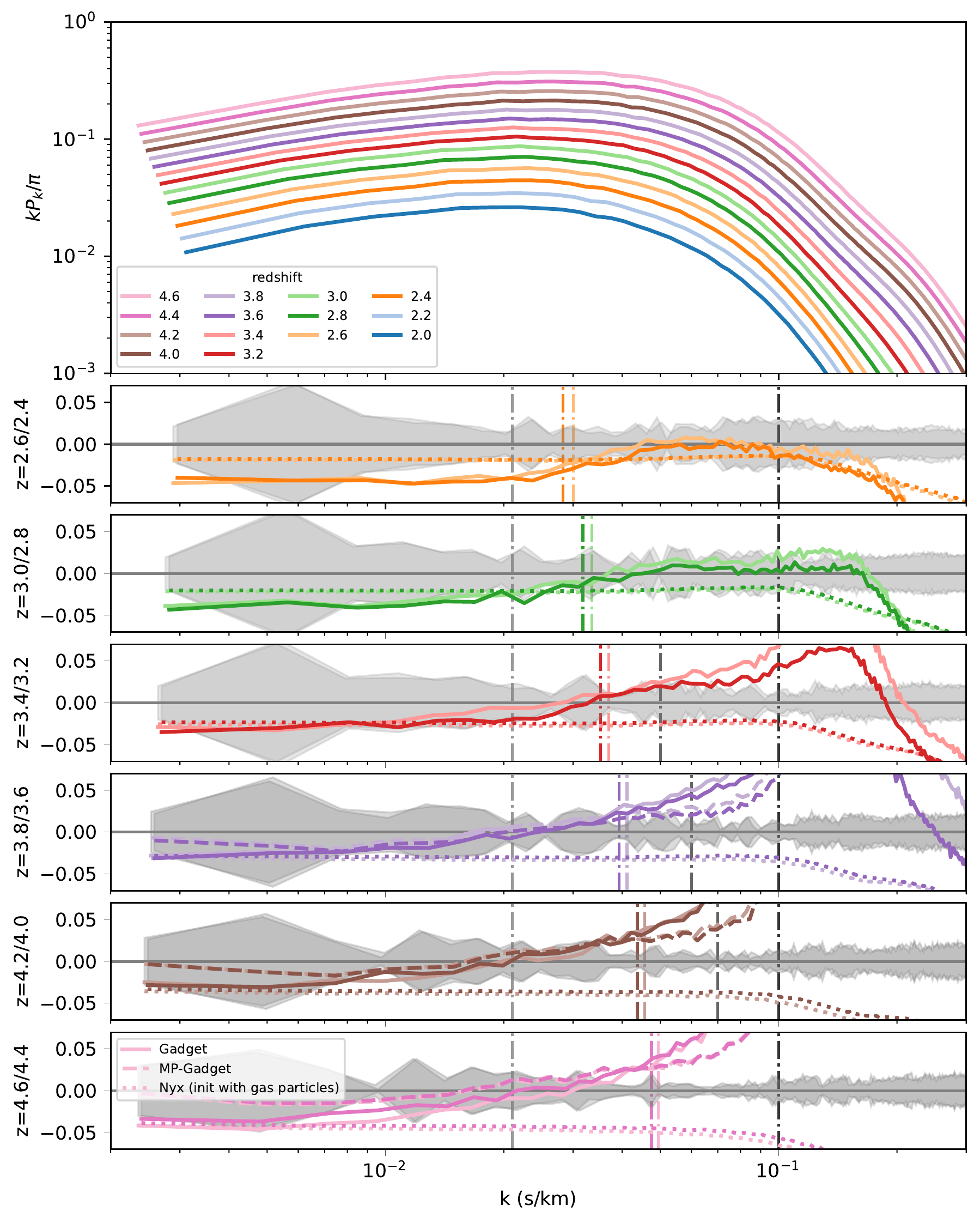}
    \caption{Top: Flux power spectrum for our fiducial Nyx simulation at our output redshifts (colors).
    Below: Relative differences $P/P_\mathrm{Nyx}-1$ of Gadget (solid) and  MP-Gadget (dashed) to Nyx runs. For Nyx we're also showing an alternative simulation using the exact same initial conditions as for Gadget (dotted, see main text). Grey ranges show an estimate of {sample} variance computed from skewers along the 3 different box dimensions. Vertical bars show the largest $k$ covered by measurements from (e)BOSS~\cite{Palanque-Delabrouille:2013,Chabanier:2019} (light grey), X-SHOOTER~\cite{Yeche:2017} (grey) \& UVES/HIRES (dark grey, typical cut to avoid noise and metal contamination) observations and estimates for the DESI resolution limit (colored by redshift). {Each panel shows two neighboring redshifts, illustrated with two shades of the same color as in the top plot.}}
    \label{fig:PF_Gadget_vs_Nyx}
\end{figure}

The matter power spectrum is not directly observable from \lya{} forest data, and instead one uses flux statistics to constrain small scale clustering of the gas.  One of most sensitive such statistics is the flux power spectrum.
Ideally, one would observe the full 3d flux power in a similar way as for power spectra based on tracer populations like galaxies or quasars.  However, this has proven to be hard in case of \lya{} forest data where angular separations between sightlines are far larger then the line-of-sight separations between pixels in a spectrum \cite{McDonald:2003,Arinyo-i-Prats:2015,Font-Ribera:2018}.
Still, even the 1D power spectrum, i.e. the line-of-sight power in individual sightlines averaged over many lines of sight, provides significant cosmological information. 
This is true especially on small scales, where 1D power spectrum is one of the prime statistical quantities for cosmological constraints \cite{Chabanier:2019,palanque-delabrouille2019}.
We want to emphasize that this statistics does not easily allow to disentangle redshift space effects, e.g. thermal line broadening, from the underlying cosmological signals, leading to a very high requirement on measurement precision and modeling accuracy.
Previous comparisons between different simulation approaches~\cite{regan2007} quoted $\sim 5\%$ differences in the \lya{} forest power, but were done on simulations of lower resolutions.

Even though we used the same UVB rates in all simulations, they do not recover the same mean flux. We thus rescaled the optical depth along our lines of sight to obtain identical evolution in all our simulations -- a very common approach in \lya{} forest studies.
Specifically, we rescaled mean fluxes to the redshift evolution of Ref.~\cite{Palanque-Delabrouille:2015a} which is in reasonable agreement with Ref.~\cite{Becker:2013}:
\begin{equation}
\bar{F} = \langle e^{-\tau}\rangle = e^{-\tau_{\rm eff}} \;\; {\rm where} \;\; \tau_{\rm eff} = 2.53\times 10^{-3} \, (1+z)^{3.7}.
\end{equation}

We show the flux power spectrum of our fiducial simulations in \autoref{fig:PF_Gadget_vs_Nyx}. We can see that all three codes agree on the $<5\%$ level on large scales (small k), but larger deviations are visible at high redshifts ($z\gtrsim 3.4$) for small scales. We can also see that agreement between Nyx and MP-Gadget (where available) is generally somewhat better than for Nyx and Gadget-3. {Note that part of the differences might be sourced in Nyx and Gadget having been run with different force resolutions. Due to the mean flux matching applied in our analysis such differences can affect all scales. However, we would not expect force resolution to account for differences seen at high redshift bins due to the \lya{} forest signal originating predominantly from underdense regions.}

On large scales (small k) and especially for high redshifts the differences between both simulations can in parts be explained by the different way of initializing the simulations.
To check the effect this difference has on the flux power we also show results from a Nyx run where we initialize the baryon grid from the SPH particles created in the Gadget initial conditions.
We can see that this decreases the flux power by up to $4\%$ on all observable scales (and additionally leads to a cutoff feature for very small scales).
On the largest scales probed by our simulations this brings Gadget and Nyx in percent level agreement at $z\ge3.2$ (note that similarly this also improved agreement on the matter power, see \autoref{sec:sim-matter-dist-cmp}).
Note again, that generating initial conditions for Baryons in typical cosmological initialization codes consistently is a non-trivial problem with accurate solutions only having been developed very recently (e.g.~\cite{Bird:2020, Hahn:2020}). We therefore leave more careful analyses of the impact of initial conditions to future work.

The source of the small scale differences is less clear, there are multiple effects that could play a role.
One the one hand there are clearly numerical effects as at least for high redshifts the differences between SPH codes and Nyx reach a similar level as Nyx's resolution convergence (see \autoref{sec:requirements}). We did not attempt to run (MP-)Gadget simulations of even higher resolution for resolution convergence tests due to computational limitations.
On the other hand, differences in thermal evolution could be a source of differences. However, these differences are very small and are larger for smaller redshifts (see \autoref{sec:sim-thermal-cmp}). Thus, those effects cannot explain the visible variation in the small scale power.
Finally, we tested the impact of very overdense gas on the \lya{} forest (see \autoref{sec:density-dependence}) as for this gas differences in star formation model or resolution would be most extreme, but found that, especially at high redshifts, strongly overdense gas does not affect the \lya{} forest in either code.

In the end we conclude that few percent differences in the flux power between Gadget and Nyx based simulations are generated. 
However, some of those differences would be marginalized over as they mimic changes of {model parameters. For example, over the range of k accessible with DESI, the low-redshift difference could be absorbed by a $\sim 3$~\% shift in mean transmitted flux, and the high-redshift difference could be mimicked by a $\sim 20$~\% change in the IGM temperature (varied through the UV heating rate, see Eqn~\ref{eqn:heat}). In that case, the physical meaning of those parameters would therefore be lost to some degree.} A full characterization of differences between these codes is beyond the scope of this work (see e.g.~\cite{embersonBorgCubeSimulation2019,Villasenor:2020} for other recent works containing code comparisons in the \lya{} forest regime). 
Given the different treatment of simulation runs regarding initialization, hydro method and analysis toolchains, one might treat the result as a  worst-case scenario and small differences as seen here are expected.
We are thus reassured about the robustness of the \lya{} forest power as a tracer of small-scale distribution of matter. 
The two codes produce results of comparable accuracy for a given boxsize and resolution. We will use Nyx in the remainder of this work due to its better efficiency with respect to computation time. 

\section{Resolution and Boxsize convergence with Nyx in light of DESI requirements} \label{sec:requirements}

\begin{figure}
    \centering
    \includegraphics[width=0.66\textwidth]{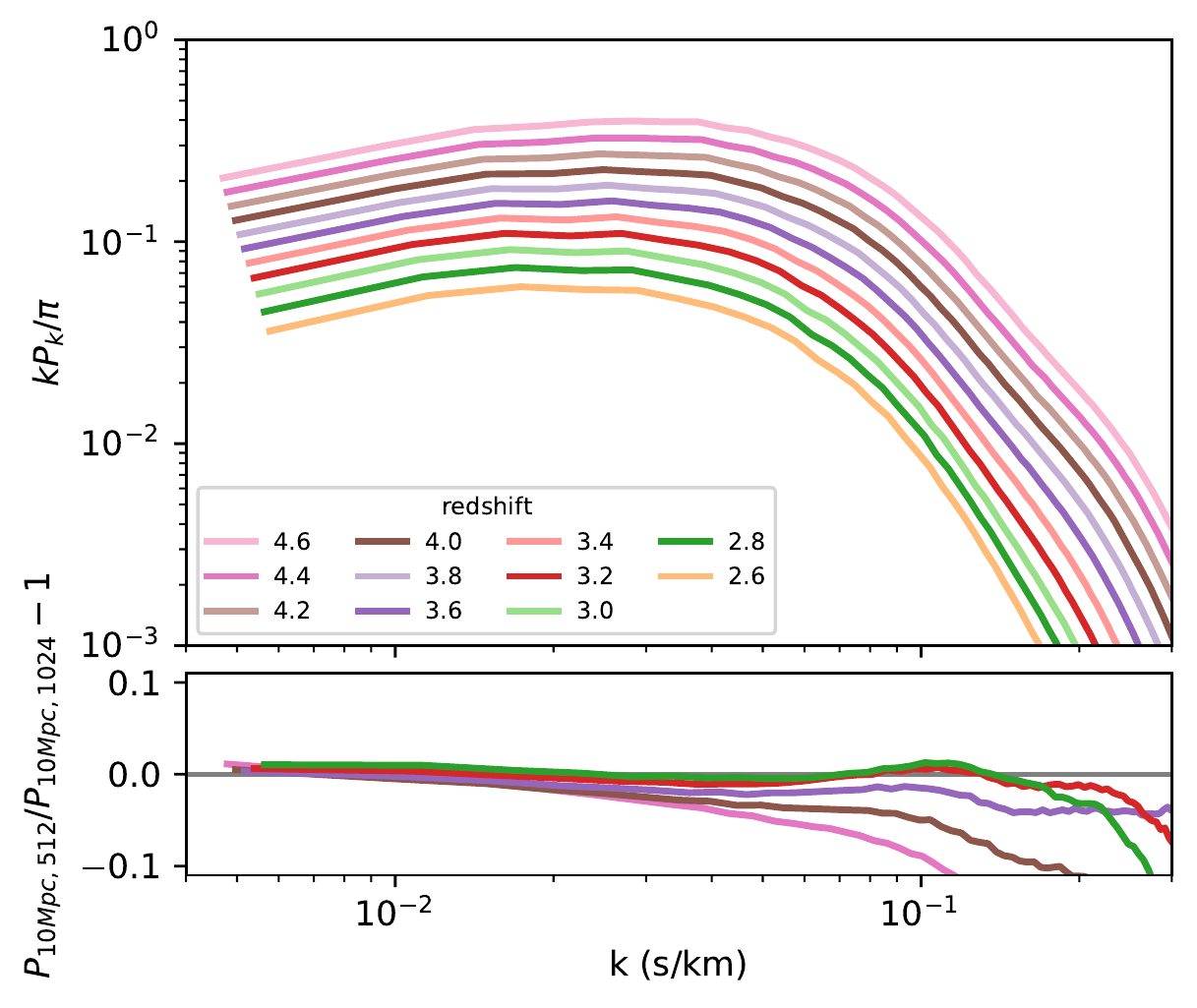}
    \caption{Top: The P(k) estimated from a $15\,\mathrm{Mpc}$ simulation with $1024^3$ resolution elements taking lines of sight along the x, y and z axis. 
    Bottom: Relative difference between simulations with $512^3$ and $1024^3$ resolution elements.}
    \label{fig:Nyx_resolution}
\end{figure}

\begin{figure}
    \centering
    \includegraphics[width=0.9\textwidth]{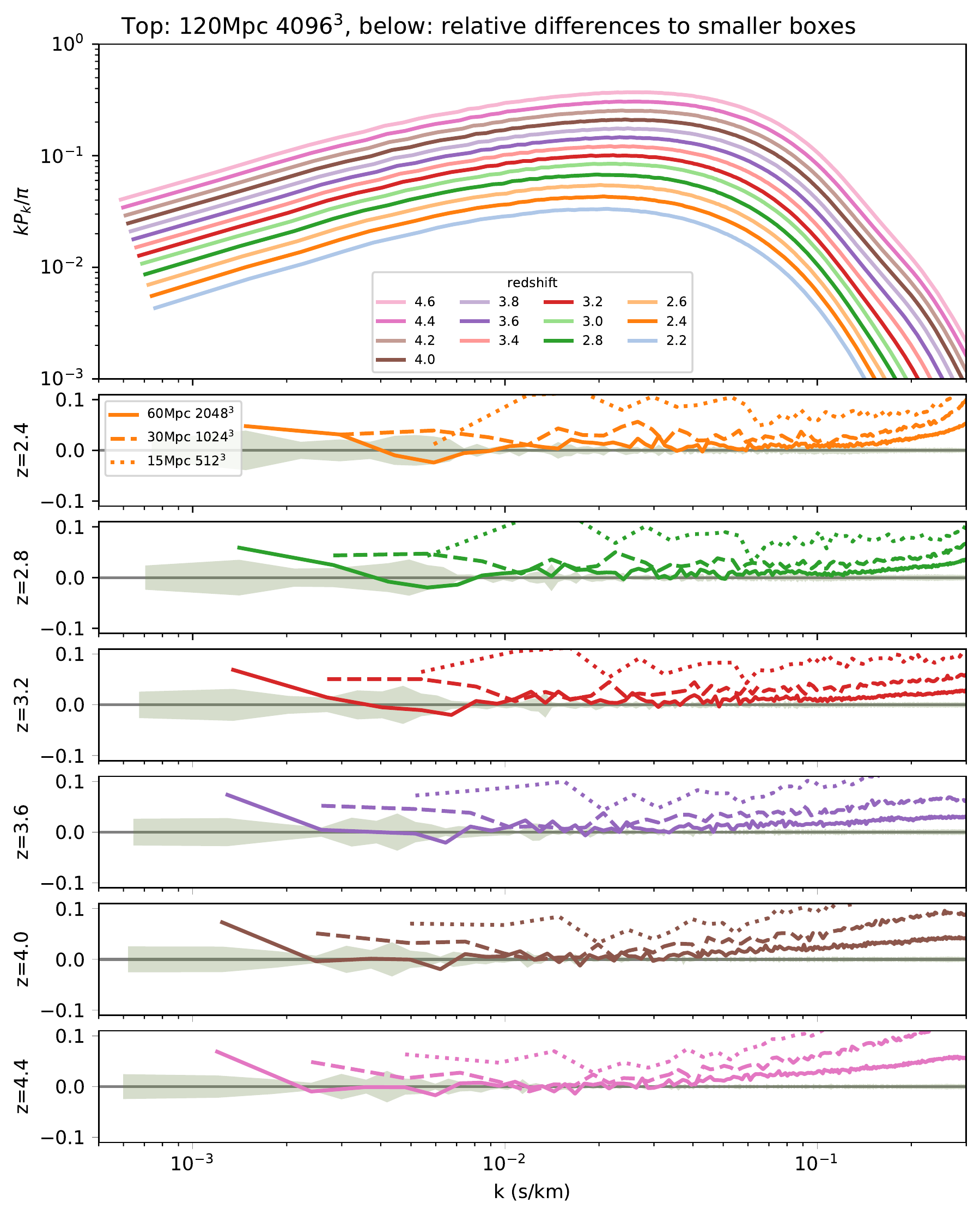}
    \caption{Top: The P(k) estimated from a $120\,\mathrm{Mpc}$ simulation with $4096^3$ resolution elements taking lines of sight along the x, y and z axis. 
    Below: Relative difference with smaller simulation volumes at different redshifts. The gray shaded area shows the standard deviation between the measurements for the 3 different line-of-sight orientations as a rough estimate for {sample} variance in the 120 Mpc box. Note that its impact especially on the largest scale mode inside the box is significant.
    }
    \label{fig:Nyx_boxsize}
\end{figure}

We show convergence tests of Nyx regarding resolution and boxsize in Figures \ref{fig:Nyx_resolution} and \ref{fig:Nyx_boxsize} respectively. Note that for both cases we average the power measured for skewers along all three simulation directions. {For the line-of-sight power spectrum, this is equivalent to using three independent simulation boxes and allows us to reduce {sample} variance accordingly. }

Regarding resolution, we are showing a comparison of our fiducial runs with to twice as resolved simulations of same box size. We can clearly see that for our main scales of interest ($k<0.05\,\mathrm{s/km}$) resolution convergence is better than $5\%$ at all redshifts. 
For $z\leq 3.2$, i.e. for the redshift range where the most data are available or anticipated, resolution convergence is $\lesssim 1\%$ for the whole range of scales $k<0.1\,\mathrm{s/km}$ probed by existing observations\footnote{Note that for $k>0.05 \mathrm{s/km}$, available data relies on observations with high-resolution spectrographs leading to significantly smaller datasets ($\mathcal{O}(1000)$ spectra vs. $\mathcal{O}(1000000)$) and thus measurement precision. At the smallest scales ($k>0.1$) also observational effects, e.g. metal absorption systems along the line of sight or accuracy of resolution or noise corrections in the measurements, additionally complicate observations. Similarly, the highest redshift data is limited by the abundance of sufficiently bright background quasars, see e.g.~\cite{Palanque:2016}, thus constraining the achievable precision. For high $k$ and high $z$ therefore a lower modeling accuracy is acceptable.}.
The observed decrease in convergence with redshift is expected and can be explained by a decrease in mean transmission inside the simulation especially at higher redshifts~\cite{Lukic:2015} due to the lack in resolution.

For boxsize convergence, we note that simulation boxes of 120Mpc achieve sub-percent level convergence over most of the scales relevant for the \lya{} forest. We can clearly see that while for scales of $k\sim 0.01 \mathrm{s/km}$ and redshifts $z>3.2$ even our 30Mpc boxes reach a few-$\%$ level of convergence, while 15Mpc boxes at low redshifts are never converged beyond the $10\%$-level.
However, for lower redshifts and especially on small and large scales significant differences due to boxsize are observable.
Boxsize affects the simulation in multiple ways.
On the largest scales even with the 120Mpc simulations there is a $~2\%$ {sample} variance (gray band in figure) effect which affects progressively smaller scales and increases in amplitude as boxsize decreases.
In addition, the fundamental mode of the box is already mildly nonlinear\footnote{Note that this also depends on cosmological parameters, i.e. when generating  a grid of simulations ideally one would want to still sample linear modes for the most extreme parameters.} for the smaller box sizes~\cite{Lukic:2015} suppressing evolution of modes. Finally, lacking large scale modes lead to underestimated bulk flow and also affects shock heating~\cite{Lukic:2015}, which both would lead to increased small scale power as visible in the figure.
Note that boxes of at least this size are also required to access large scale modes ($k\sim 0.001 \mathrm{s/km}$) in any case.

We thus conclude that for analyses of DESI like datasets at least $120\,\mathrm{Mpc}$ boxes with $4096^3$ resolution elements need to be run, which is in line with previous works \cite{Borde:2013, Lukic:2015}. Note that performing such a simulation with Nyx on modern CPU-based supercomputers takes $\sim2$ million CPU-h. Therefore with a typical PRACE allocation about 15 such simulations can be run.

\section{Generating a cosmological emulator for the \lya{} forest}\label{sec:emulator}
\subsection{Adopted parameters and grid design}
To perform fits on observational \lya{} forest data, one needs to be able to predict the observables for different combinations of cosmological parameters. As for the required accuracy and typical supercomputing allocations only a limited number of simulations can be performed (see \autoref{sec:requirements}), one requires a highly accurate interpolation scheme allowing such predictions and to be able to perform a cosmological fit, nevertheless.

Traditionally, for cosmological parameter estimation from the \lya{} forest this interpolation was performed in the following way. 
First, a set of simulations was run  assuming a fiducial model, along with  additional simulations that each modified  one or two of the input parameters at a time.  
These simulations were then used to build a first-order~\cite{Viel:2006} or second-order Taylor expansion including cross-term contributions~\cite{Borde:2013}.
The Taylor-expansion approach ensures a good accuracy near the fiducial model and along the axes of the expansion, but shows a rapid loss of accuracy as one explores off-axis regions of the parameter space.

However, in recent years a promising alternative has become available. Based on initial works of \cite{heitmann2006Cosmiccalibration, habib2007CosmicCalibrationConstraints} the framework of Gaussian process emulation has been established in cosmology (e.g. \cite{Zhai:2019,Wibking:2019,McClintock:2019}) and has also been utilized in multiple analyses of the \lya{} forest \cite{Walther:2019,birdEmulatorLymanalphaForest2018,Rogers:2020}.
Here, we follow this general approach. 
We generate a grid of simulations with parameters $\Theta_\text{sim}$ arranged in a space filling Latin-hypercube design (LHD).
We then train a Gaussian process to emulate each of the modes for new sets of parameters $\Theta_\text{emu}$.
The main advantages of this approach are a more uniform interpolation uncertainty which does not rely on a single fiducial model, the possibility to evaluate said uncertainty on-the-fly without running additional simulations, and the option to increase accuracy in a certain region of parameter space by running additional simulations.
Especially doing the latter on a Taylor expansion scheme can require re-running the complete grid around a new fiducial model of interest.
Furthermore, potentially large interpolation uncertainties in specific regions of the parameter space might stay unnoticed, as uncertainties are not estimated in the interpolation process.
Additional simulations would be needed to test the interpolation accuracy at chosen points of the parameter space. 
The simulations are parametrized by the cosmological parameters $\Theta_\text{sim}=\{A_\mathrm{p}, n_s,$ $ \Omega_m, H_0\}$ where $n_s$, $\Omega_m $ and $H_0$ are the scalar spectral index, matter density and Hubble parameter, respectively, and $A_\mathrm{p}$ is a rescaled spectral amplitude, $A_\mathrm{p} = A_s [2\pi/(1.6\times10^{-2})]^{n_s-1}$, following~\cite{birdEmulatorLymanalphaForest2018}, with $A_s$  the conventionally-defined spectral amplitude.
Note, however, that the initial matter power spectrum depends on $\omega_m=\Omega_m h^2$, but is independent of $h$ for simulations at fixed $\omega_m$.
Hence lower parameter degeneracies are to be expected if analyses are performed in a parametrization that uses $\omega_m$ and not $\Omega_m$.

For this emulator we additionally rescaled the UV heating rates (a typical approach when performing \lya{} forest analyses (e.g. \cite{Viel:2004,Becker:2011}) to obtain different thermal histories and thus different amounts of pressure smoothing. The heating rates have been constructed as:
\begin{equation}\label{eqn:heat}
    \epsilon = A_\mathrm{UVB} \Delta^{B_\mathrm{UVB}} \epsilon_\mathrm{fid}
\end{equation}
with $\epsilon_\mathrm{fid}$ being the fiducial heating rates and $\Delta$ being the overdensity. 

To vary the thermal state of the IGM we followed a two-fold approach. First, to generate different thermal histories we rescaled UV heating rates by a factor $A_\mathrm{UVB}$ which effectively changes the temperature at mean density $T_{0,\mathrm{sim}}$ at every redshift of the simulation, while keeping its density dependence (characterized by a logarithmic slope $\gamma_\mathrm{sim}-1$) fixed. 

In addition, to modify the instantaneous temperature-density relation which e.g. affects Doppler broadening we use a post-processing procedure at every output redshift: we rescale temperatures of all hydro cells by a factor $T_0/T_{0,\mathrm{sim}}\Delta^{\gamma-\gamma_\mathrm{sim}}$, such that a new instantaneous temperature-density relation $T=T_0 \Delta^{\gamma-1}$ is generated while pressure smoothing is unmodified, allowing for independent modification of both effects. 
While this approach will produce non-negligible artifacts for the smallest scales and is thus not usable for an interpretation of \lya{} forest data from high-resolution datasets, the effects of this approach on large scales  $k\lesssim 0.04\,\mathrm{s km^{-1}}$ as measured e.g. in the DESI survey are small \cite{takhtaganov:2019}. 

Finally, we generate different mean transmissions $\bar{F}$ in the \lya{} forest by rescaling the optical depth in post-processing.
For each redshift the values of $\Theta_\text{post}=\left\{T_0, \gamma, \bar{F}\right\}$ were chosen from a second LHD, i.e. we generated one LHD covering the 5 simulation parameters and one LHD for each output redshift covering the 3 post-processing parameters.

\begin{figure}
    \centering
    \includegraphics[trim=0.1in 0.1in 1.95in 1.95in, clip,width=0.64\textwidth]{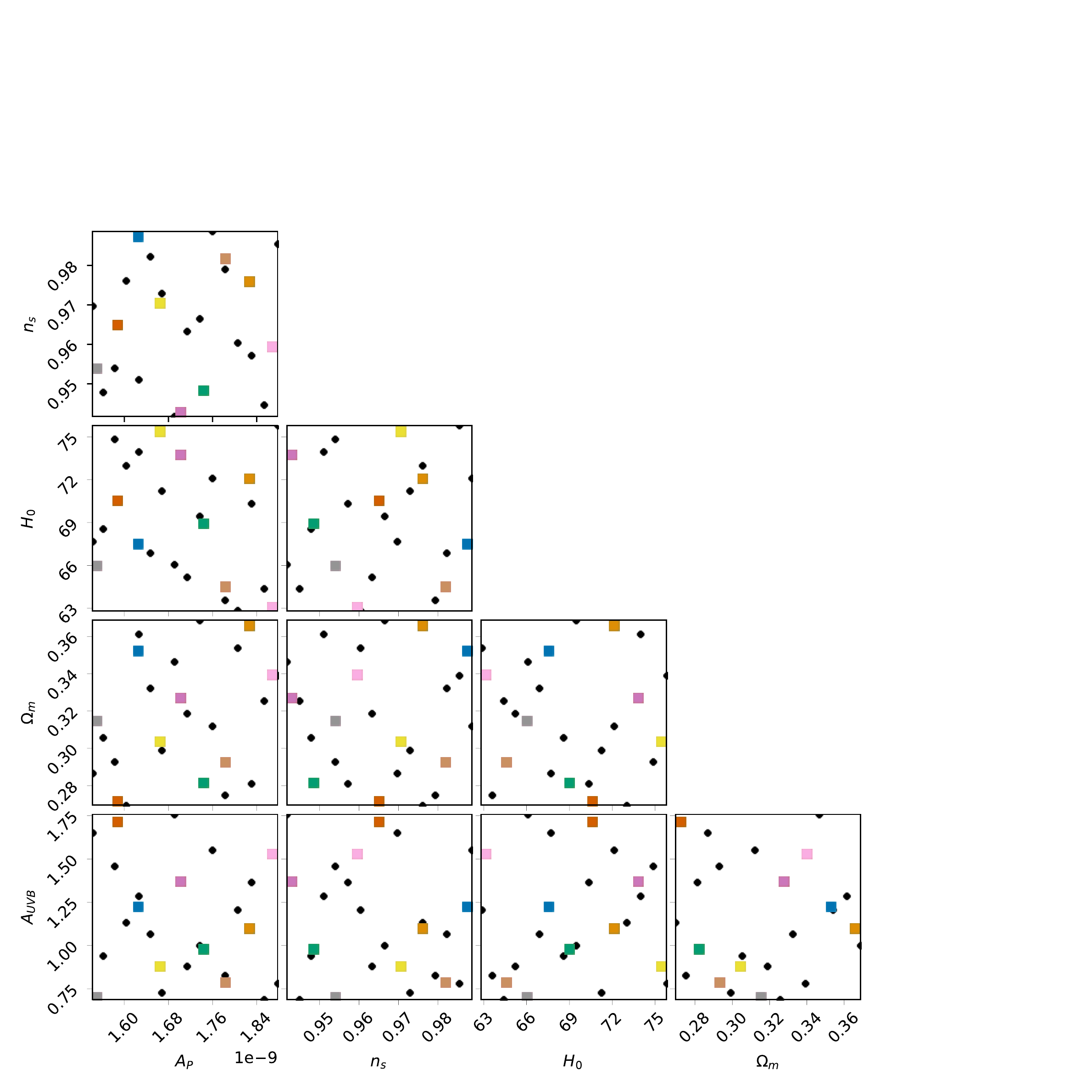}
    \includegraphics[trim=1.95in 0 0.1 0.1,clip,width=0.35\textwidth]{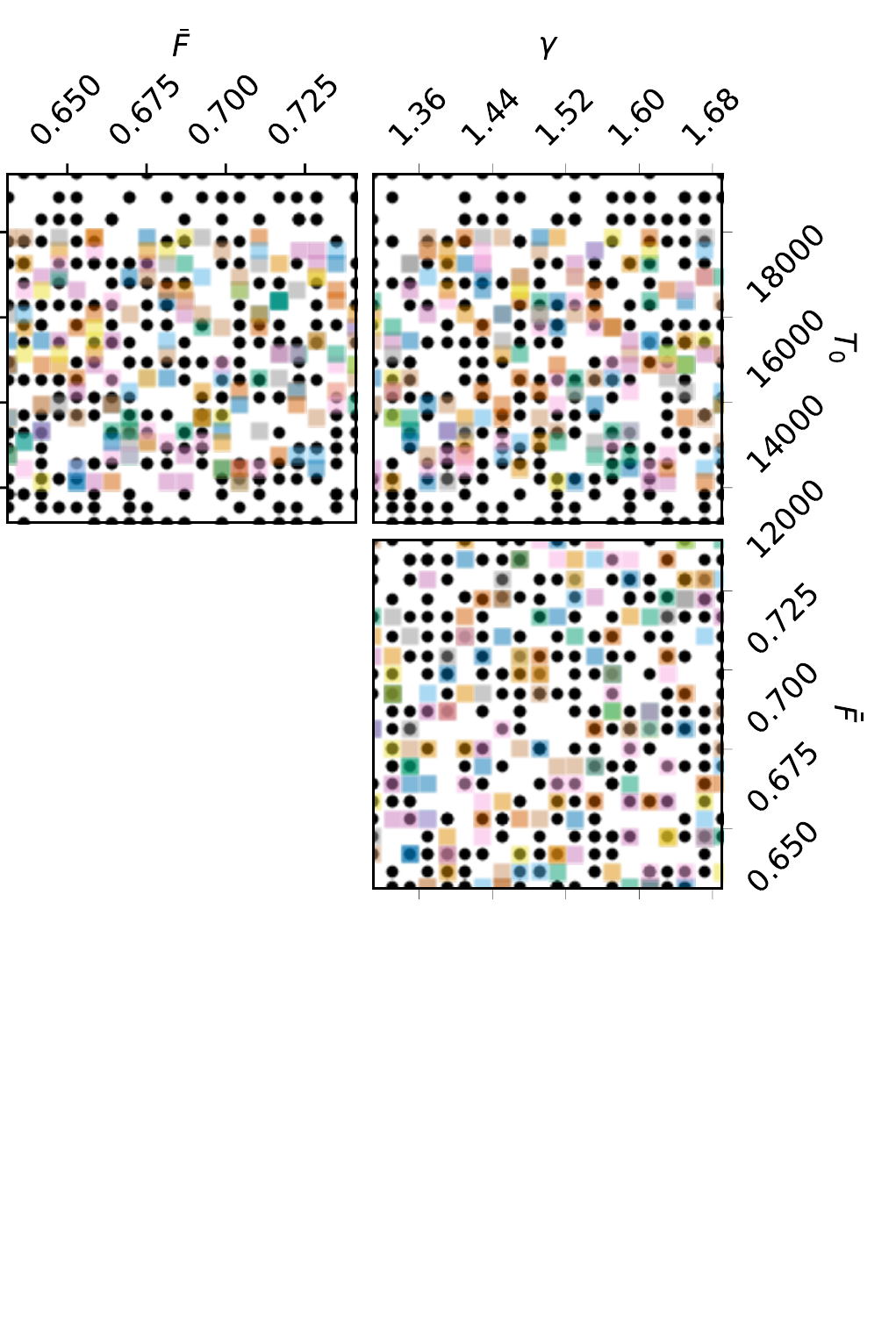}
    \caption{Grids in simulation input (left) and post-processed thermal parameters for z=3 (right), black points show training sample, colored points show testing grid, note that the thermal parameters are actually consisting of a different LHD for each set of cosmological parameters at each redshift.}
    \label{fig:emu_grids}
\end{figure}

The Gaussian process emulator is then built individually for each redshift using the combined set of $\Theta_\text{sim}$ and $\Theta_\text{post}$ as input parameters. The cosmological parameters are centered on their best-fit value from~\cite{planckcollaborationPlanck2018Results2018a} and span a $\pm 5~\sigma$ range around these values. 
The simulation parameters of the training grid as well as the post-processing parameter grid for z=3 are listed in \autoref{tab:lh_params} and shown in \autoref{fig:emu_grids} as black points.

In order to verify our approach and quantify its precision, we additionally generated an independent Latin hypercube (called "Testing" in \autoref{tab:lh_params} and illustrated using colored points in \autoref{fig:emu_grids}) for cosmological parameters, which was otherwise treated in the same way. We can then use our emulation procedure at the parameter values of the testing grid and compare to the output of a post-processed simulation.   To test an end-to-end analysis (see \autoref{sec:endtoend}), we also generated "consistent" thermal evolutions for those mocks, i.e. the model realizations were generated to follow (broken) power laws in $T_0$, $\gamma$ and $\bar{F}$.

\begin{table}
    \caption{Input parameters for the simulations that were used to build and verify the emulation scheme}
    \label{tab:lh_params}
    \centering
    \begin{tabular}{ccccccccc}
        \hline\hline
        ~ & $\Omega_\mathrm{m}$ & $h$ & $10^9 A_\mathrm{s}$ & $n_s$ & $A_\mathrm{UVB}$ & $\sigma_8$ & $10^9 A_\mathrm{p}$& $\Omega_m h^2$\\ \hline
        Training 0  &0.3186 & 0.6523 & 2.1325 & 0.9633 & 0.8839 & 0.7929 & 1.7131 & 0.1356\\
        Training 1  &0.3537 & 0.6283 & 2.2897 & 0.9602 & 1.2081 & 0.8316 & 1.8055 & 0.1396\\
        Training 2  &0.3120 & 0.7209 & 1.8823 & 0.9886 & 1.5512 & 0.8782 & 1.7587 & 0.1621\\
        Training 3  &0.3612 & 0.7391 & 2.1790 & 0.9509 & 1.2860 & 1.0646 & 1.6255 & 0.1973\\
        Training 4  &0.2870 & 0.6772 & 1.8495 & 0.9696 & 1.6513 & 0.7293 & 1.5424 & 0.1316\\
        Training 5  &0.2930 & 0.7484 & 2.0838 & 0.9540 & 1.4573 & 0.9267 & 1.5834 & 0.1641\\
        Training 6  &0.3688 & 0.6943 & 2.1207 & 0.9665 & 1.0016 & 0.9724 & 1.7358 & 0.1778\\
        Training 7  &0.2992 & 0.7119 & 1.9638 & 0.9727 & 0.7328 & 0.8468 & 1.6687 & 0.1517\\
        Training 8  &0.3322 & 0.6688 & 1.8313 & 0.9822 & 1.0662 & 0.7966 & 1.6470 & 0.1486\\
        Training 9  &0.2695 & 0.7299 & 1.8527 & 0.9759 & 1.1349 & 0.7951 & 1.6043 & 0.1436\\
        Training 10 &0.2810 & 0.7031 & 2.3633 & 0.9571 & 1.3690 & 0.8623 & 1.8293 & 0.1389\\
        Training 11 &0.3056 & 0.6857 & 2.1337 & 0.9479 & 0.9409 & 0.8329 & 1.5628 & 0.1437\\
        Training 12 &0.2752 & 0.6362 & 2.0193 & 0.9791 & 0.8303 & 0.6588 & 1.7819 & 0.1114\\
        Training 13 &0.3464 & 0.6605 & 2.3945 & 0.9417 & 1.7578 & 0.9049 & 1.6908 & 0.1511\\
        Training 14 &0.3253 & 0.6442 & 2.5774 & 0.9448 & 0.6884 & 0.8610 & 1.8535 & 0.1350\\
        Training 15 &0.3392 & 0.7578 & 2.0488 & 0.9854 & 0.7800 & 1.0484 & 1.8780 & 0.1948\\
        \hline
        Testing 0   &0.2820 & 0.6900 & 2.3713 & 0.9484 & 0.9843 & 0.8354 & 1.7421 & 0.1343\\
        Testing 1   &0.2927 & 0.6455 & 1.9891 & 0.9817 & 0.7882 & 0.7079 & 1.7832 & 0.1220\\
        Testing 2   &0.3038 & 0.7541 & 1.9834 & 0.9705 & 0.8808 & 0.9454 & 1.6627 & 0.1728\\
        Testing 3   &0.3659 & 0.7214 & 2.1057 & 0.9761 & 1.1000 & 1.0277 & 1.8253 & 0.1904\\
        Testing 4   &0.3272 & 0.7376 & 2.3932 & 0.9429 & 1.3737 & 1.0407 & 1.7019 & 0.1780\\
        Testing 5   &0.3153 & 0.6600 & 2.0423 & 0.9539 & 0.7053 & 0.7831 & 1.5503 & 0.1373\\
        Testing 6   &0.3525 & 0.6748 & 1.7515 & 0.9874 & 1.2293 & 0.8261 & 1.6243 & 0.1605\\
        Testing 7   &0.2717 & 0.7055 & 1.9570 & 0.9649 & 1.7156 & 0.7714 & 1.5869 & 0.1353\\
        Testing 8   &0.3396 & 0.6313 & 2.3818 & 0.9594 & 1.5352 & 0.8297 & 1.8684 & 0.1354\\
        \hline        
        Refine 0    &0.3030 & 0.6600 & 2.3852 & 0.9510 & 1.0000 & 0.8201&&\\
        Refine 1    &0.3418 & 0.6400 & 2.2490 & 0.9430 & 1.0500 & 0.8244&&\\
        Refine 2    &0.3345 & 0.6400 & 2.3195 & 0.9460 & 1.0250 & 0.8248&&\\
        \hline\hline
    \end{tabular}\\
\end{table}

\subsection{Dynamic Range and Box size for the Emulator Test}
Running a full hydrodynamical simulation grid fulfilling the requirements shown in \autoref{sec:requirements} is too costly for the purpose of verifying the emulation procedure. Therefore we opted to adopt the so-called splicing technique introduced by~\cite{McDonald:2003} and extensively used in the simulation grids of~\cite{Borde:2013, Rossi:2014} to generate a high-dynamic range output power spectrum. This method uses a set of three simulations, one with a large box (in this case $60\,\text{Mpc}$) but relatively low resolution ($N_\mathrm{fluid}=512^3$), one smaller high-resolution box ($15\,\text{Mpc}$, $N_\mathrm{fluid}=512^3$) and one that matches the size of the small box and resolution of the large box (i.e. $15\,\text{Mpc}$, $N_\mathrm{fluid}=128^3$) allowing to compute correction factors for the lack of mode-mode-coupling in the small box and lack of resolution in the large box. 
The power spectrum is eventually computed for three different regimes:
\newcommand{\smallbox}{15}
\newcommand{\largebox}{60}
\newcommand{\highres}{512}
\newcommand{\lowreslarge}{512}
\newcommand{\lowressmall}{128}
\begin{equation}
P_F(k)= 
\begin{cases}
    P_\text{F,\largebox,\lowreslarge}(k)\times \frac{P_\text{F,\smallbox,\highres}(k_\text{min,\smallbox})}{P_\text{F,\smallbox,\lowressmall}(k_\text{min,\smallbox})}           & k \leq k_\text{min,\smallbox}\\
    P_\text{F,\smallbox,\highres}(k)\times \frac{P_\text{F,\largebox,\lowreslarge}(k_\text{Nyq,\largebox,\lowreslarge}/4)}{P_\text{F,\smallbox,\lowressmall}(k_\text{Nyq,\largebox,\lowreslarge}/4)}            & k \geq k_\text{Nyq,\largebox,\lowreslarge}/4\\
    P_\text{F,\largebox,\lowreslarge}(k)\times \frac{P_\text{F,\smallbox,\highres}(k)}{P_\text{F,\smallbox,\lowressmall}(k)}              & \text{elsewise} \\
\end{cases}
\end{equation}
where $k_{\text{min},L}$ denotes the fundamental mode, $k_{\text{Nyq},L,N}$ is the Nyquist frequency, and $P_{\text{F},L,N}$ is the power taken from the box of length $L$ with $N^3$ fluid elements.
While the splicing technique causes biases that need to be corrected in cosmological analyses, see e.g.~\cite{Borde:2013,Lukic:2015,Palanque-Delabrouille:2015b}, this approach is sufficient for the purpose of testing the emulator as we are only interested in relative changes of the power with cosmological parameters. Note, that we use the splicing approach for creating the emulator, but also to generate our mock observations, i.e. for the purposes of this test splicing influences our ``data'' in the same way as the model. 

For our simulations we  use constant box sizes in units of $\mathrm{Mpc}$ and not $\mathrm{Mpc}\; h^{-1}$ as suggested by~\cite{Sanchez:2020} for high-redshift analyses to avoid artificially introducing a $h$-dependence. At our redshifts of interest, the linear matter power spectrum is nearly independent of $h$, and the expansion history (which directly affects hydrodynamical evolution, ionization, and thermal history) is solely dependent on $\omega_m$ ($h$ only impacts it via the dark energy contribution, which is still small even at $z=2$):
\begin{equation}
H(z)=H_0 \sqrt{\Omega_m (1+z)^3 + (1-\Omega_m)} = \sqrt{\omega_m (1+z)^3+(h^2-\omega_m)} \times  100 \mathrm{km\, s^{-1} Mpc^{-1}}.
\end{equation}

In addition, the conversion between length-scale-based power spectra $P_l$ (as obtained from simulations) and  velocity-based power spectra $P_v$ (as produced by \lya{} forest observations) depends on $H(z)$ as:
\begin{equation}
    P_v=\frac{H(z)}{1+z} P_l
\end{equation}
For fixed $\omega_m$, there is therefore only a very mild dependence of this conversion on $h$ at the redshifts of interest, which would be increased with simulations using constant sizes in $\mathrm{Mpc}/h$ units\footnote{Note that for fixed $\Omega_m$ the conversion indeed is independent of $h$ when using $\mathrm{Mpc}/h$ units, however in this case the initial power depends on $h$ which is undesirable.}.

\subsection{Emulating the Power Spectrum}
To build an emulator of the \lya{} forest power spectrum we proceed by training a Gaussian process using the spliced power spectrum outputs of our LHD. We also tried training the Gaussian process on a principle component decomposition of the output power spectra. Note that this approach requires recombining the individual PCA components and thus leads to covariant error estimates produced by the interpolation routine as well as a mildly increased overall uncertainty. Therefore we chose to emulate modes separately for the moment.

For the Gaussian process we use a stationary Matern-5/2 covariance function:
\begin{align}
    K_m(r) &= \sigma_0^2\left(1+\sqrt{5 r^2} + \frac{5}{3} r^2 \right) \exp\left(-\sqrt{5}r\right)\\
    r^2 &= (\vec{x}_i - \vec{x}_j)^T C (\vec{x}_i - \vec{x}_j)\\
    C &= \mathrm{diag} (l_{1}^{-2} \cdots l_{n}^{-2})
\end{align}
with an independent length scale $l_{i}$ for each parameter $\theta_{i} \in \Theta$. We also tested our approach replacing $K_m$ with a squared exponential function, but this produced higher emulation errors at the positions of the testing grid.

Thus, $K$ has the hyper-parameters $\{l_{i}\}$, $\sigma_0$ which were optimized for each mode and redshift bin independently. Typical length scales obtained fall between 3 and 20 times the range covered by simulations for each parameter. 
The Gaussian process was generated using the python package \texttt{George}~\cite{Ambikasaran:2014}.

\subsection{Emulation Errors}

\begin{figure}
    \centering
    \includegraphics[width=\columnwidth]{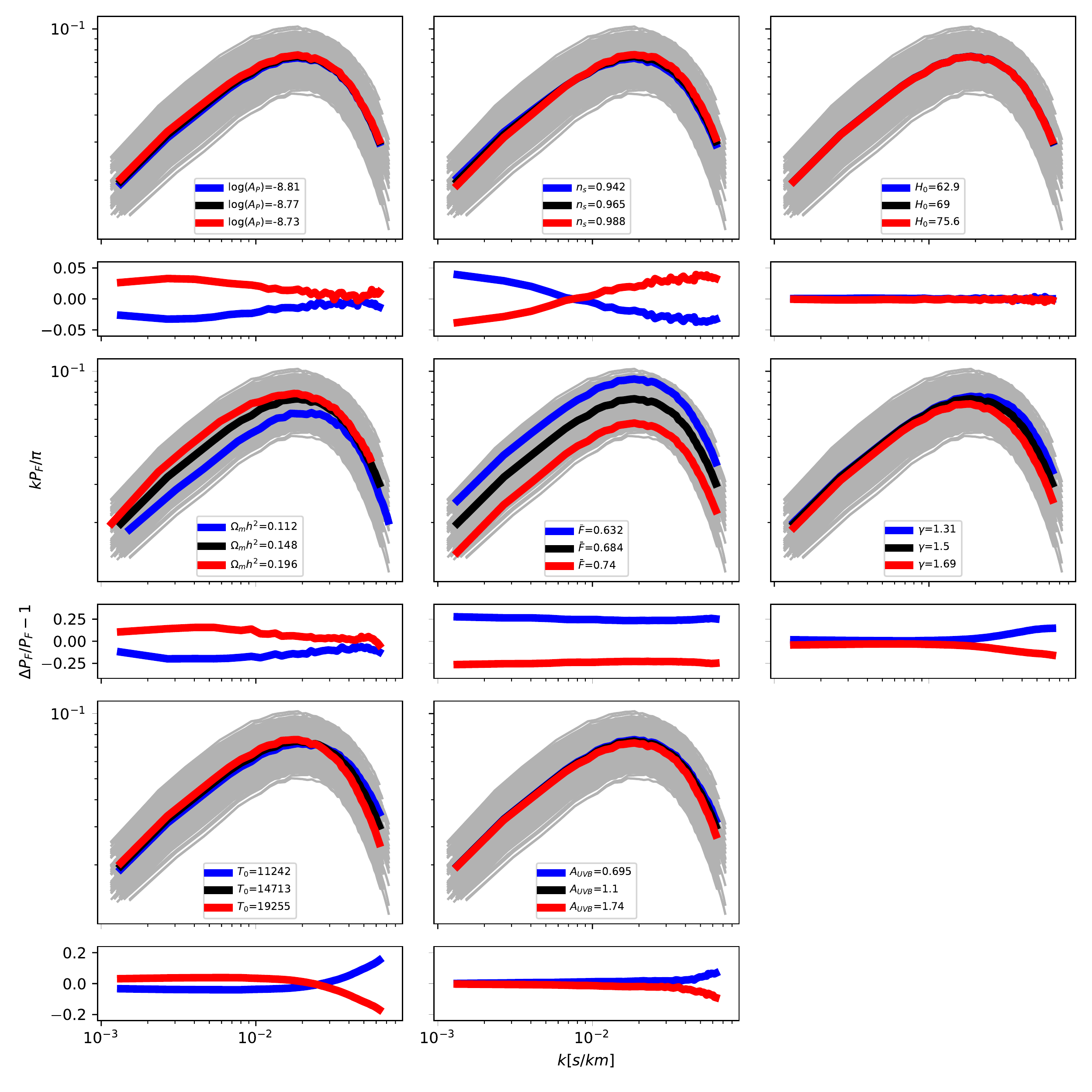}
    \caption{Effect of changes in cosmological and astrophysical parameters on the power spectrum at z=3 keeping other parameters fixed (in absolute scale or relative to the power spectrum for the central value of the parameter). Parameter values are chosen at the 1, 50 and 99 percent of the grid extent. All other parameters are kept to the grid center. Note that while those variations are redshift dependent in detail, overall trends are similar.}
    \label{fig:parameter_changes}
\end{figure}

\begin{figure} 
    \centering
    \includegraphics[width=\textwidth]{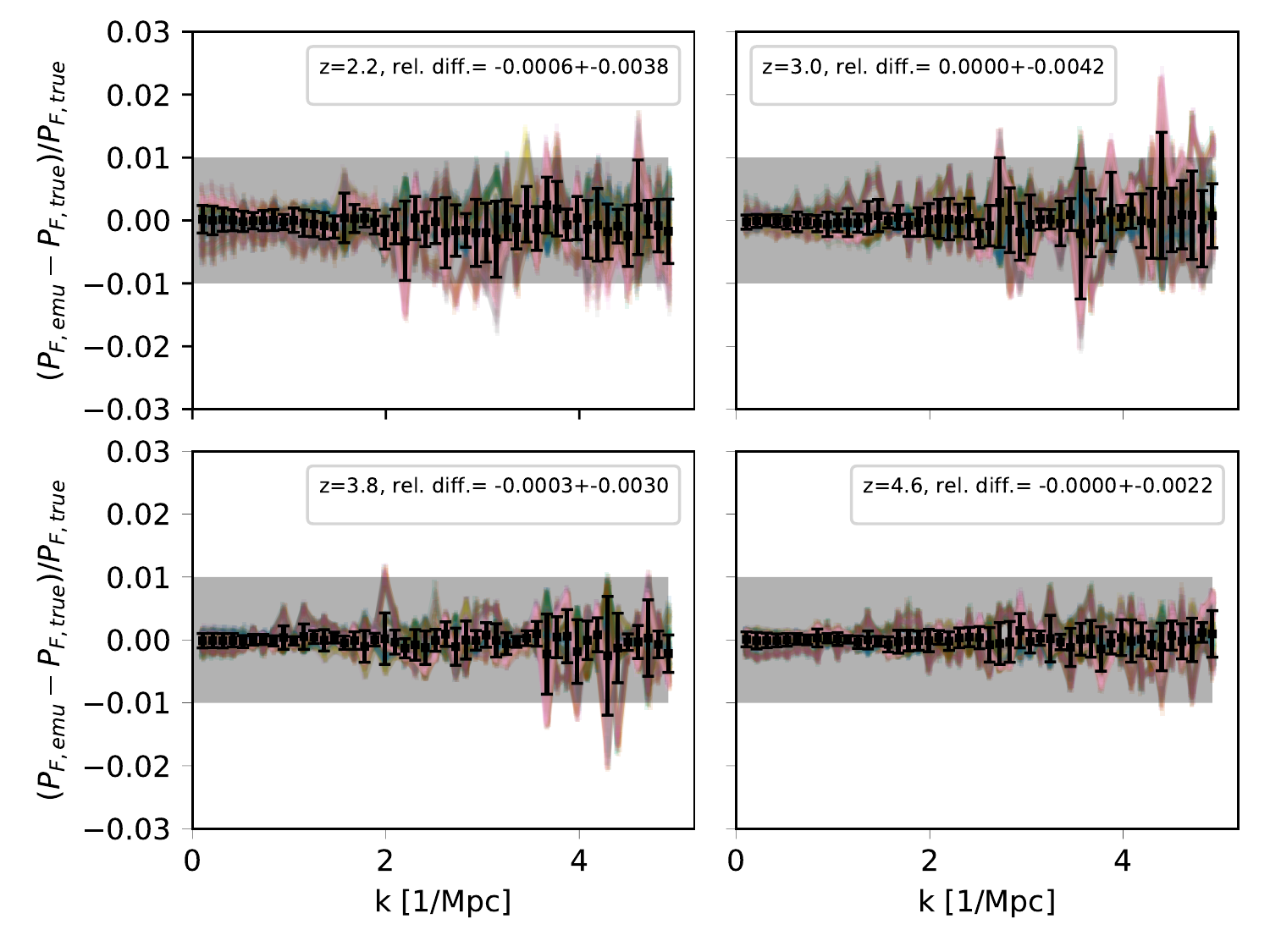}
    \caption{Accuracy of power spectrum emulation for different redshifts. Shown is the relative difference between the power spectra from our testing grid and from the emulator at the same parameter values. Note that training and testing are performed on  completely separate Latin-hypercube grids.
    Model colors are matched with \autoref{fig:emu_grids}. Different lines of the same color correspond to different post-processings of the same cosmological model. {Black squares and error bars show the average difference across all models (as an estimate of bias) and the standard deviation between models for each mode (as a systematic uncertainty of the emulation). The grey area shows a $1\%$ region as an estimate of DESI expected uncertainty.}
   }
    \label{fig:interp_accuracy}
\end{figure}

To test the accuracy of our emulation procedure we used an additional set of 9 "testing" simulations which cover the same range in parameters as our training grid, but consist of a different Latin hypercube realization (cf. \autoref{tab:lh_params}). For each of these simulations we again generated a grid of post-processed parameters.

Using our emulator we first check its dependency on the different cosmological and thermal parameters involved. This is shown in \autoref{fig:parameter_changes} where we compare the power spectra obtained from the emulator when varying each parameter within it's simulated range while fixing all parameters to their central values within the grid. We can clearly see that e.g. varying the Hubble parameter $H_0$ by $10\%$ with $\Omega_m h^2$  kept fixed only affects the flux power at the 2 per mille level. For all other parameters, we obtain roughly the expected behavior: changing $n_s$  mostly changes the slope of the flux power,  changing $\bar{F}$ only changes the overall amplitude,  increasing the temperature (via post-processed $T_0$ or $\gamma$ or via increasing the $A_{UVB}$ and thus increasing the pressure smoothing) modifies the cutoff scale to smaller k. We note that changing $A_\mathrm{p}$ or $\Omega_m h^2$ has a very similar effect on the flux power for this parametrization, hinting towards strong degeneracies between both parameters. Further parameter optimization is required to better decouple the parameters impact. 

To test the accuracy of our power spectrum predictions, we emulate a power spectrum for each set of input parameters in our testing grid and compare it with the  power spectrum measured from the corresponding simulation. In \autoref{fig:interp_accuracy} we can see the relative difference between both types of models for one redshift. Different colors correspond to different cosmological parameters, different lines of the same color are for the same cosmological model but different astrophysical parameters. We show the prediction error estimated by the emulation scheme as error bars. We see that the emulator creates a negligible overall bias, but introduces a $\sim 0.2 \text{--} 0.4\%$ uncertainty on the prediction. 
In addition to the emulation errors of each model, we  show the mean interpolation error as {black points} and the scatter in interpolation errors as its error bars. The $1\%$ uncertainty expected from a DESI-like measurement is shown as a gray band. We can see that while individual interpolation errors at particular positions of the grid exceed the $1\%$ level especially for large k,  the emulator is able to produce unbiased predictions of the power with better than $0.4\%$ accuracy overall.

\subsection{Fitting cosmological parameters based on our testing grid}
\subsubsection{General fitting procedure}\label{sec:fitting_procedure}
To test the accuracy of our emulation approach when inferring cosmological parameters from future datasets such as the recently-started DESI survey~\cite{DESIFinalDesignReport}, we performed an end-to-end fitting test based on mock data.
All fits are performed using a Gaussian likelihood:
\begin{align}
    \mathcal{L}(\theta) &\propto \exp \left(-\frac{1}{2} \sum_{z\in \{z_i\}} \Delta(\theta,z) \mathrm{C(z)}^{-1} \Delta(\theta,z)\right)\\
    \mathbf{\Delta}(\theta) &= \mathbf{P}_\mathrm{data}-\mathbf{P}_\mathrm{emulator}(\theta)
\end{align}
where $C$, $\mathbf{P}_\mathrm{emulator}(\theta)$ and $\mathbf{P}_\mathrm{data}$ are, respectively, the data covariance matrix for a given redshift bin which we chose to be diagonal for this test, emulated power spectrum vector for some parameter combination $\theta$, and mock data power spectrum vector in a single redshift bin $z_i$. For simplicity, we assume measurement k-bins to be uncorrelated, and assume the diagonal covariance elements to either follow the errors of the latest eBOSS measurements~\cite{Chabanier:2019}, or to represent $1\%$ errors on each mode with $k<5\, \mathrm{Mpc}^{-1}$ at all redshifts, which would be beyond the reach even of upcoming datasets.

The parameters $\theta$ for our fit consist of the four cosmological parameters $\omega_m$, $h$, $A_\mathrm{p}$, $n_s$, of one parameter regarding a global rescaling of the UV-background heating rate $A_\mathrm{UVB}$, and of 7 parameters characterizing the evolution of the thermal and ionization state of the IGM using our post-processing approach assuming a double power-law evolution for $T_0(z)$, and a power law evolution for each $\tau_\mathrm{eff}(z)$ and $\gamma(z)$.

Our emulator was trained in the same way as in the previous section. All tests were performed using the emulator  to predict the 1D power spectrum at the parameters of one of the "testing" simulations (which we will call  the "true" simulation later). We did not include any measurement error. While for most tests we used our set of "training" simulations to build the emulator, we also performed tests using different setups, generating the emulator by:
\begin{itemize}
    \item using the whole set of "training" and "testing" models excluding the "true" simulation to assess the impact of additional "randomly placed" simulations on cosmological fits,
    \item adding the "true" simulation for either of those setups to assess the approach in the best-case scenario where interpolation error at the true parameters is zero, i.e. where the true model should actually be the maximum likelihood model. Note that even in this setup, the phase space around the truth will still need to be explored and does have non-zero emulation error.
    \item adding models to the default emulator chosen close to peaks in the posterior distribution of prior fits in order to check how refinement simulations could improve the analysis.
\end{itemize}
In all cases, a separate emulator was built for each redshift: we did not attempt to build a common emulator for all redshifts.

We performed our inference using the affine invariant MCMC sampler \texttt{emcee}\cite{ForemanMackay:2013} and adopting flat priors for each parameter within its simulated range.

\subsubsection{Fitting a fiducial model in different configurations} \label{sec:single-model-fitting}
\begin{figure}
    \centering
    \includegraphics[width=\linewidth]{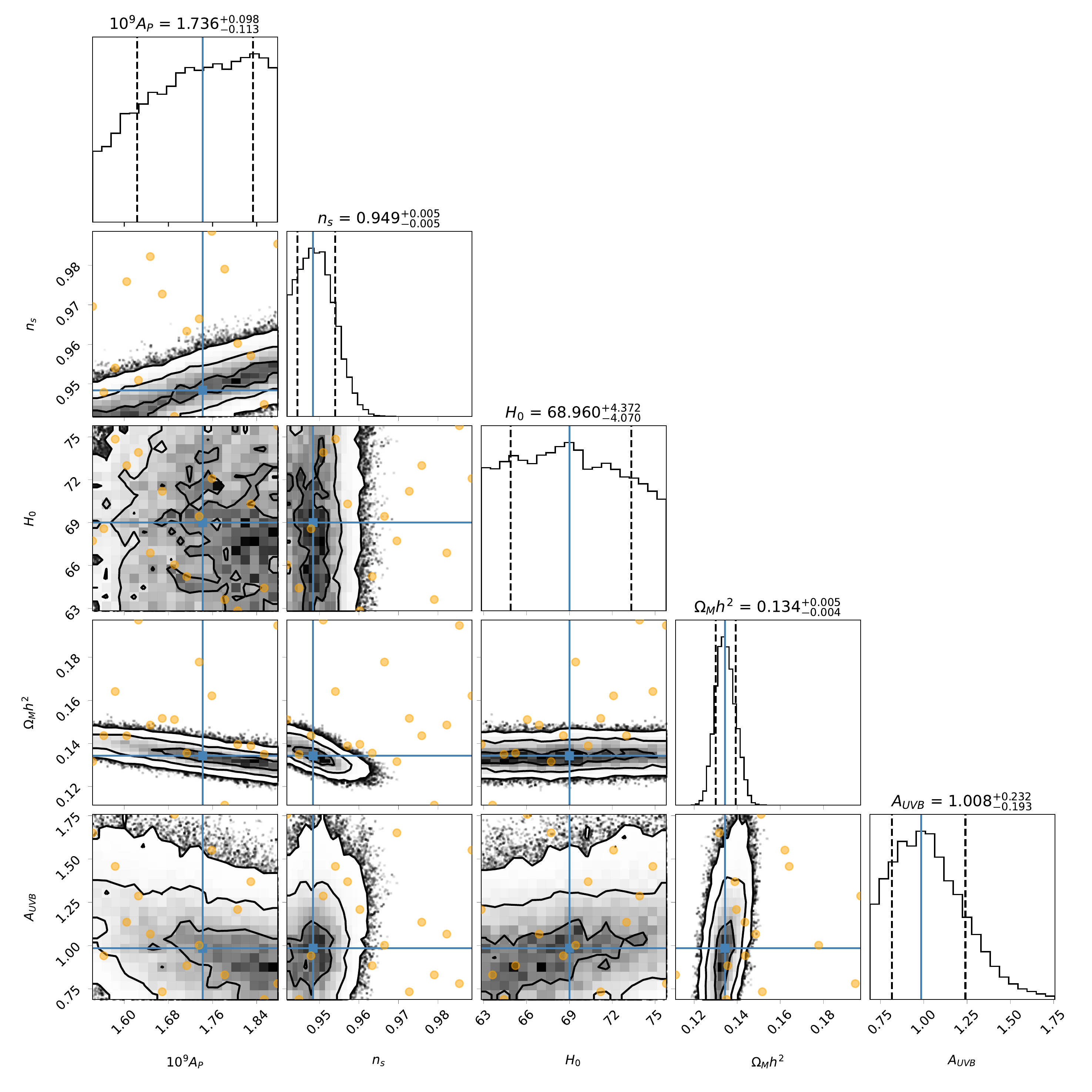}
    \caption{Fit of fiducial model using eBOSS like errors. The true model position is shown in blue, yellow points show the models in our training grid. Contours show $1, 2, 3\sigma$ confidence regions, dashed lines in the histograms correspond to the $68\%$ confidence interval. We can see that the contours match the true value in all parameters. From this analysis the eBOSS Lya flux power alone should be able to constrain $\Omega_m h^2$ to less than $2\%$ as well as $n_s$ to $\sim 0.5\%$. The Hubble parameter, however, stays nearly unconstrained as expected.}
    \label{fig:eboss_errors}
\end{figure}

In this section, we choose Testing 0 as our fiducial simulation. We fit this fiducial model assuming different scenarios for the measurement errors, and with different configurations of the emulator.

We first use our standard emulator and assuming eBOSS-like measurement errors. Corner plots showing the resulting posterior distributions are shown in Figure \ref{fig:eboss_errors}. We obtain $0.5\%$ constraints on $n_\mathrm{s}$ (similar to the $0.6\%$ obtained in~\cite{palanque-delabrouille2019} for a fit to the eBOSS \lya{} data alone) and $3.7\%$ constraints on $\Omega_m h^2$ (a bit larger than the constraints on $\Omega_\mathrm{m}$ for the same fit in~\cite{palanque-delabrouille2019}\footnote{but note that there are correlations between $\Omega_\mathrm{m}$ and $h$, and that our $H_0$ constraint is weaker than the prior assumed in~\cite{palanque-delabrouille2019}}). The fits are centered on the true value as expected. The errors introduced by performing the emulation are thus expected to be subdominant for this analysis.

\begin{figure}
    \centering
    \includegraphics[width=\linewidth]{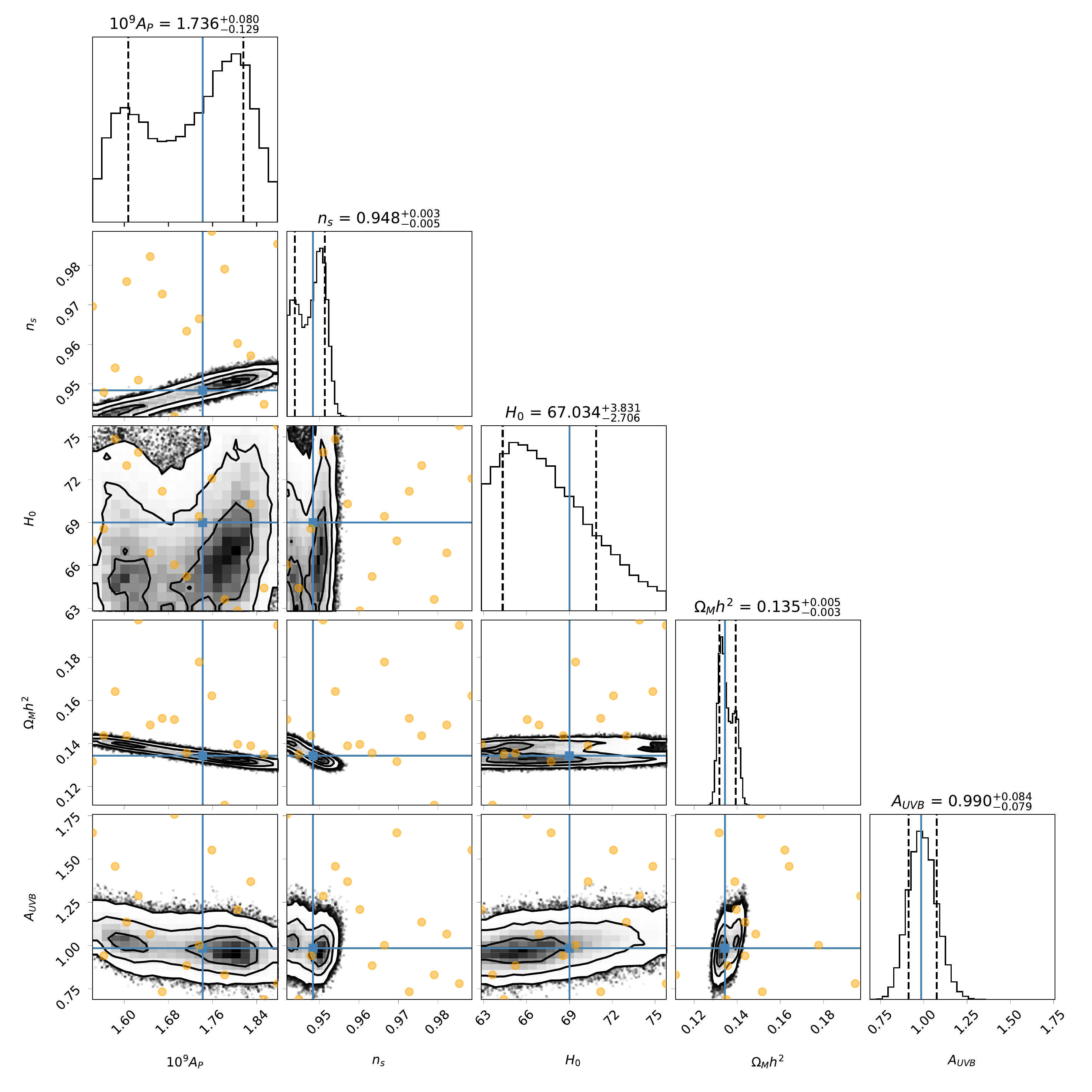}
    \caption{Same as \autoref{fig:eboss_errors}, but assuming 1\% errors.
    In this case the emulation was not producing good results close to the true value, leading to bimodalities in the posterior distribution.}
    \label{fig:bimodality}
\end{figure}
\begin{figure}
    \centering
    \includegraphics[width=\linewidth]{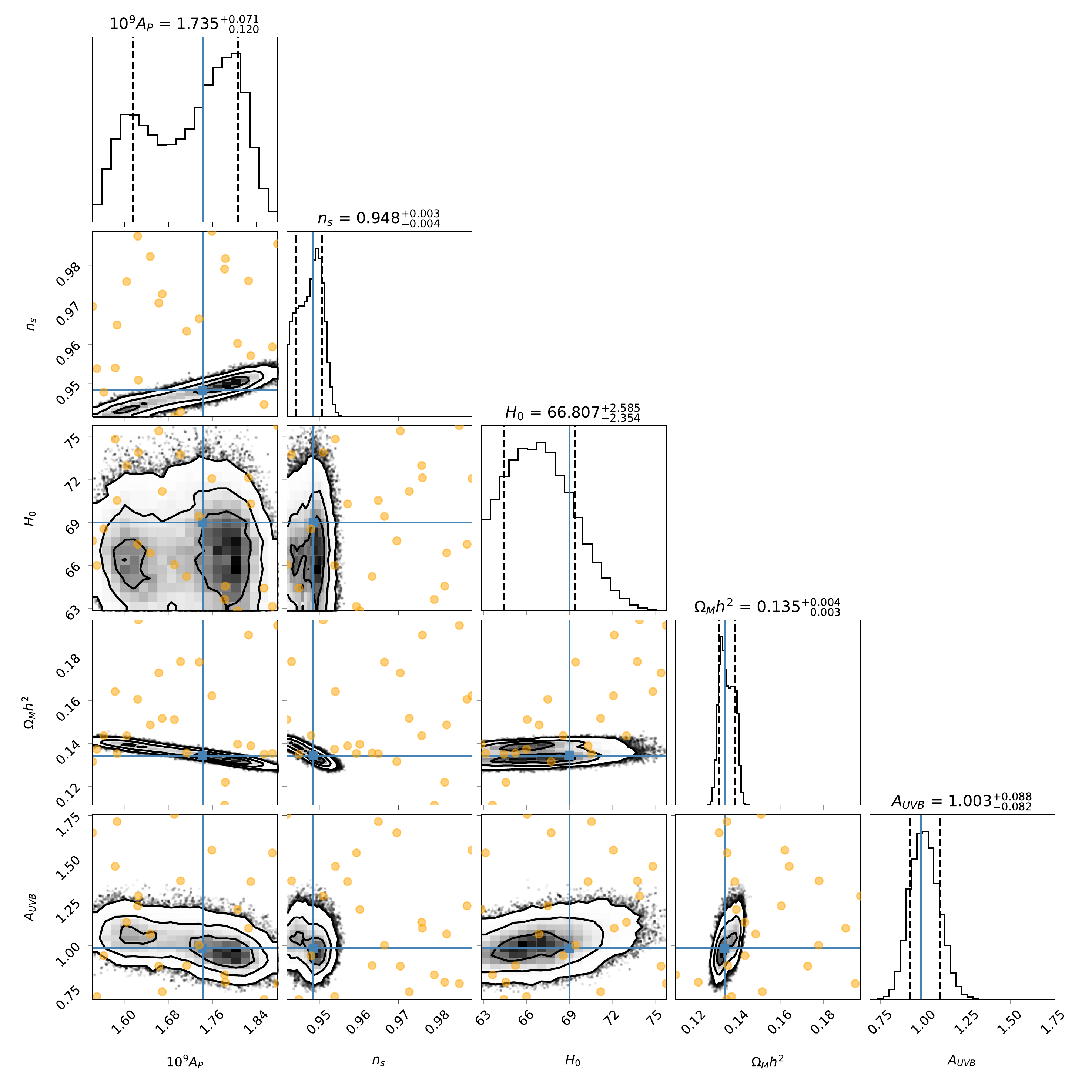}
    \caption{Same as \autoref{fig:bimodality}, but now using all testing models except for the truth for training. Note that fitting results are nearly unchanged.}
    \label{fig:with_more_training}
\end{figure}
We then assume 1\% measurement errors, more comparable to what upcoming surveys are aiming for. Note that we use $1\%$ at each redshift, however, which is better than what can be achieved in the near future with DESI for instance. Resulting corner plots are shown in Figure~\ref{fig:bimodality}.
While the observed contours shrink due to the reduced uncertainties, we also observe a bimodality  due to lacking accuracy in the emulation. 

To improve the emulation accuracy, we use the alternative training sets described in \autoref{sec:fitting_procedure}. 
When increasing the number of models in the emulator by including the whole testing grid except for the truth (i.e. increasing the number of models within our parameter phase space from 16 to 24), we obtain the contours shown in \autoref{fig:with_more_training}. Despite mild changes, the main feature of a bimodal fit remains. Therefore, even a significant increase in the number of models used to train the emulator does not improve the emulation accuracy to the level required if the additional models are placed in random positions (since the "Testing" models were also drawn along a Latin hypercube)  away from the true position (since the design is space filling and we explicitly excluded "Testing 0"). 

\begin{figure}
    \centering
    \includegraphics[width=\linewidth]{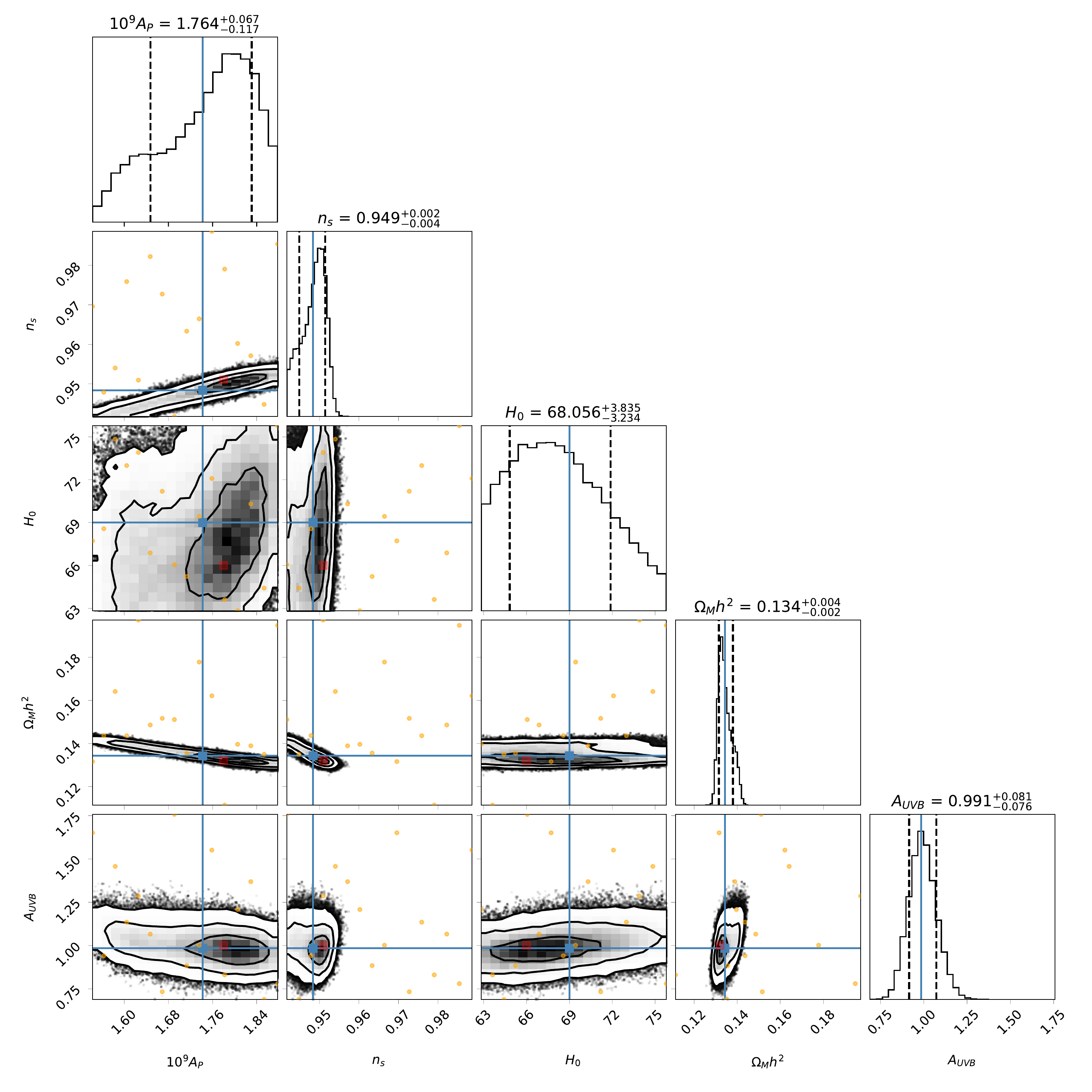}
    \caption{Same as \autoref{fig:bimodality}, but now using an additional refinement model in the training close to the posterior peak in \autoref{fig:bimodality} with {higher} $A_\mathrm{p}$ (highlighted in red).}
    \label{fig:with_truth}
\end{figure}
\begin{figure}
    \centering
    \includegraphics[width=\linewidth]{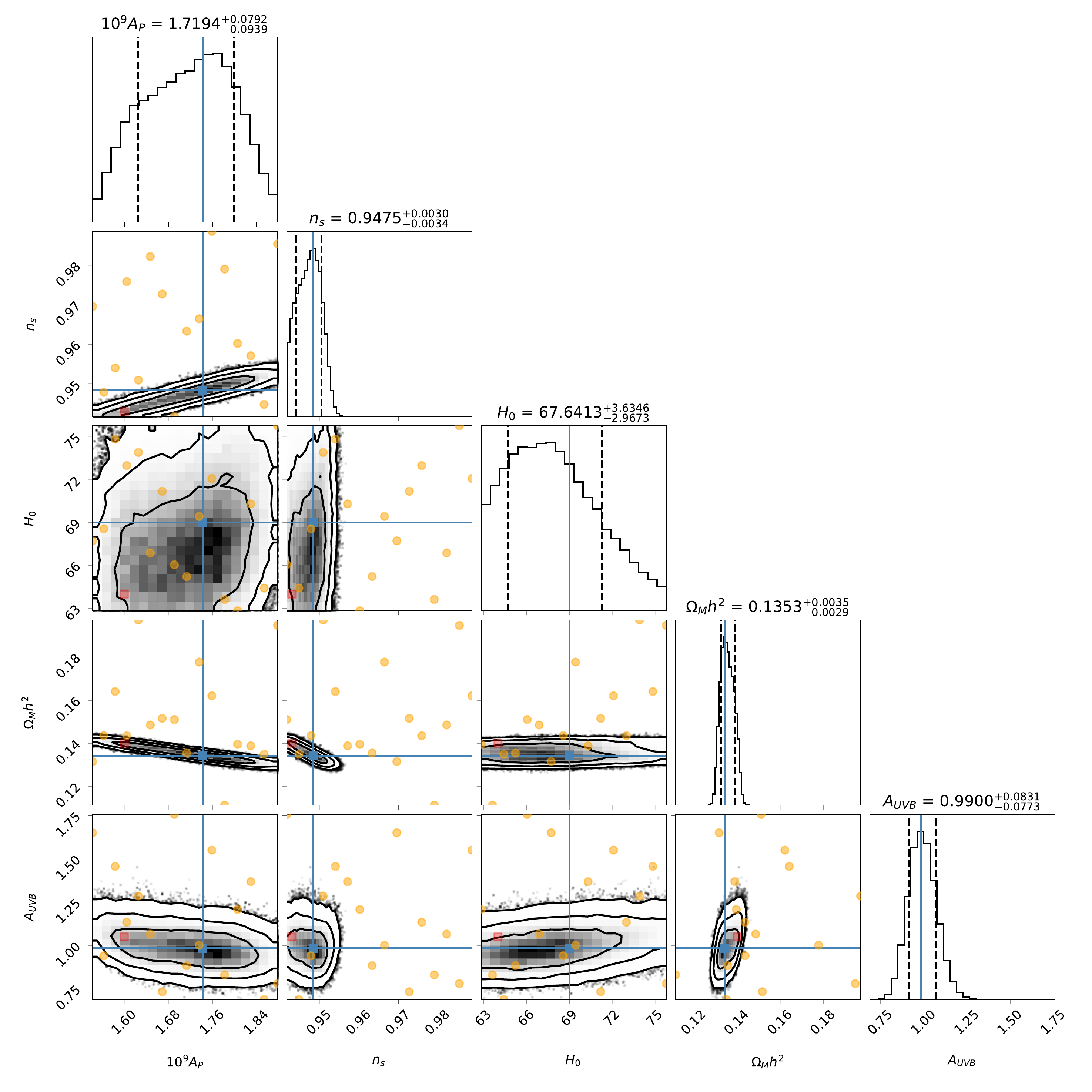}
    \caption{Same as \autoref{fig:bimodality}, but now using an additional refinement model in the training close to the posterior peak in \autoref{fig:bimodality} with lower $A_\mathrm{p}$ (highlighted in red). Note that the true model is {now well recovered} in the fit.}
    \label{fig:with_refinement}
\end{figure}
As a second test, we  tested the opposite approach, and only added the true model to our initial emulator. In this configuration, the emulation at the true parameters is perfect. However, this is not the case away from the truth, e.g. at the peaks in the posterior surface. In this configuration, the bimodality seen in the previous fits completely vanishes and the MCMC can recover the true parameters. 

Finally, we tested a possible refinement scheme: we recompute the emulator with additional models in regions of high posterior probability, i.e. models with parameters close to the peaks of the distributions shown in \autoref{fig:bimodality}. The results of the fit when refining at the peak closer to the truth are shown in \autoref{fig:with_truth}, results when refining at the peak further away are in \autoref{fig:with_refinement}. 
We can clearly see that in both cases the previous bimodality has vanished.
Comparing the contours in \autoref{fig:with_truth} and \autoref{fig:with_refinement}, we can see that while there are differences at the $1\sigma$ level, the $2\sigma$ regions are very similar. 
Therefore, additional emulator accuracy close to the likelihood peaks provided by a single additional model is more successful than including 8 additional models at random positions.
With either of the refined grids, we obtain a  precision of $0.32\%$ in $n_s$ and $2.4\%$ in $\Omega_m h^2$, i.e. roughly 2/3 of the uncertainties we obtained with the eBOSS-like error bars. 

In principle, one could also  further refine the model grid based on the posterior until full convergence is reached. A Bayesian optimization approach as described in e.g.~\cite{rogersBayesianEmulatorOptimisation2018, takhtaganov:2019} which additionally also takes into account the emulation error returned by the emulation scheme should allow such an iterative approach. While it is beyond the scope of this work to study this in more detail, we do plan to use such a technique for future analyses. Additionally, to avoid degeneracies and reduce the risk of erroneously multimodal distributions, we plan to switch to a different parametrization regarding $n_s$ and $A_\mathrm{p}$ that is better suited for the scales of interest for the \lya{} forest. We note also that for application to actual data, additional physical effects that neglected in this work (e.g. for feedback, inhomogeneous reionization) will have to be taken into account.
There are also recent developments regarding Baryonic ICs that we have not studied here \cite{Bird:2020,Hahn:2020}.

\subsubsection{Overall statistics of the fitting tests}
\label{sec:endtoend}
\begin{figure}
    \centering
    \includegraphics[width=\textwidth]{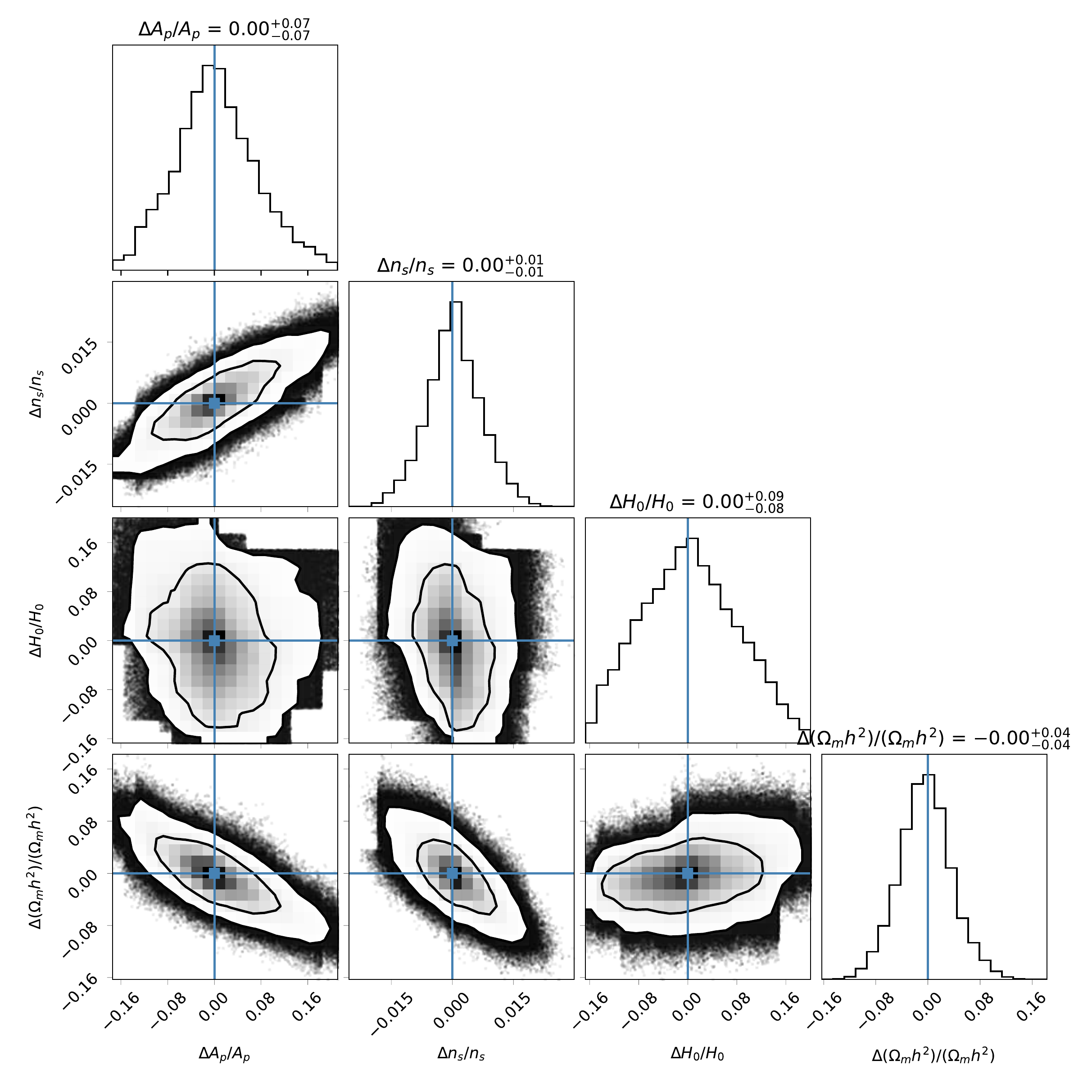}
    \caption{Merging chains for fits with each of the 9 cosmological parameter combinations based on eBOSS error bars. Shown are relative differences between fitting results and the respective truth of each model (i.e. contours are scatter within the fits and not the error on the mean for different models). We can see that overall, across the sampled parameter space and after marginalization over thermal parameters, fitting results based on the emulator are unbiased and similar constraints compared to \autoref{fig:eboss_errors} are obtained. 
    }
    \label{fig:endtoend}
\end{figure}
In \autoref{fig:endtoend} we show a summary of fits based on all 9 different testing simulations using eBOSS error bars. For this purpose we computed the relative difference to the true value for each sample in the MCMC chains and afterwards merged the chains of all different models. As the chains have the same length, each model carries the same weight and thus the corner plot shows the mean posterior PDF of all 9 fits. As all 9 models were generated in a space filling grid across the whole parameter space, it should be representative of the whole space.
We can see that the results of our emulator test are unbiased.
We can also see that similar uncertainties and parameter correlations as in \autoref{fig:eboss_errors} are obtained. We therefore are confident that the current emulation technique is suitable for cosmological analysis even with only 16 simulations.It can also be used for parameter inference on DESI \lya{} forest power spectrum data with the addition of an one-step refinement stage. 
The visible correlations are entirely due to the chosen parametrization. The interpolation could thus potentially be improved using less correlated sets of parameters.

\section{Conclusions and outlook}\label{sec:conclusion}

In this work we presented a study of modeling uncertainties in \lya{} forest analyses of future surveys, with a particular focus on DESI survey.  The work is based on hydrodynamical simulations and LHD+GP emulation.
We studied three main aspects: 
\begin{itemize}
    \item the impact of the simulation code used for the hydrodynamics (along with the scheme used for modeling the gas, i.e. grid based vs SPH) on the \lya{} 1d flux power,
    \item the convergence of Nyx-based simulations with respect to the simulation volume and resolution,
    \item the accuracy of a GP-based emulation technique assuming either eBOSS- or DESI-like uncertainties.
\end{itemize}

{Quite unsurprisingly given the vastly different algorithms used in both codes, we find that there are some remaining discrepancies in the predictions for the 1D flux power between the previously-used Gadget and Nyx simulations. These discrepancies} are small enough for the redshift and scale ranges that are relevant for DESI: data errors will be at least as high, and simulation uncertainties would be unimportant after marginalization over IGM astrophysical parameters (e.g. thermal evolution and reionization scenarios).

To obtain sufficiently converged simulations compared to the expected data accuracy of $\sim 1\%$, we determined (in agreement with \cite{Lukic:2015}) that boxes of $120\,\mathrm{Mpc}$ with a resolution of order $ 30\,\mathrm{kpc}$ (and thus with $\sim 4000^3$ gas cells) would be required.
This large dynamic range  sets strong requirements on computing systems, simulation codes and post-processing toolchains. For the Nyx code, such a simulation costs about 2 million CPU-h.
This high cost is a limiting factor regarding the number of simulations one can consider, thus careful statistical frameworks for parameter inference are mandated.

To perform parameter inference, one therefore relies upon an  interpolation scheme that achieves the required accuracy and precision with a  number of  simulations reduced to the level of 15 to 20. 
In this work, we  built an emulation scheme based on a set of 16 simulations with smaller boxes and have been able to achieve a statistical accuracy of $\sim 0.4\%$. 
This scheme has been used in an end-to-end inference test on mock data assuming uncertainties consistent with current datasets from the eBOSS survey. It showed that one can thus obtain unbiased parameter estimates.
However, for the   measurement accuracy of upcoming surveys like DESI, even the small emulation errors can affect the posterior, leading e.g. to erroneously bimodal distributions.
To solve this issue, we compared two approaches of increasing emulation accuracy by adding more simulations. In the first case, we increased the number of simulations throughout the entire parameter volume, in a Latin-hypercube fashion. This approach did not provide noticeable improvement. In the second case, we ran local refinements of the simulation grid in the spirit of recent works on adaptive emulators \cite{takhtaganov:2019,rogersBayesianEmulatorOptimisation2018}. This indeed allowed us to remove the erroneously bimodal distribution of the posterior and to improve the precision of the parameter inference without a noticeable increase in computing cost.
We thus conclude that even a single adaptive refinement step leads to significant improvements on the posterior.
Based on all of this, we expect that the full-scale version of the emulator scheme described in this work will reach the accuracy required for upcoming DESI-like datasets.

\acknowledgments
MW likes to thank Jean Sexton and Jose O\~norbe for fruitful discussions related to Nyx simulations as well as Nils Schöneberg and Julien Lesgourgues for discussions about CLASS.

We acknowledge PRACE for awarding us access to Joliot-Curie at GENCI@CEA, France through projects 2019204900, 2010PA4826.
This work was also granted access to the HPC resources of TGCC under the allocations A0050410586 and A0070410586 made by GENCI (Grand Equipement National de Calcul Intensif).

MW,EA,CR,NPD \& CY acknowledge support from grant ANR-16-CE31-0021. 

This research made use of NASA's Astrophysics Data System and arXiv. This research also made use of adstex (\url{https://github.com/yymao/adstex}).

Software:  \texttt{Nyx}~\cite{Almgren:2013}, \texttt{Gadget-3}~\cite{Springel:2005}, \texttt{MP-Gadget}~\cite{mpgadget-cite}, \texttt{gimlet}~\cite{Friesen:2016}, \texttt{extract}, \texttt{GenPk}~\cite{Bird:2017}, \texttt{fake\textunderscore{}spectra}, \texttt{CAMB}~\cite{Lewis:2000}, \texttt{CLASS}~\cite{lesgourges:2011}, \texttt{2lptic}~\cite{Crocce:2012}, \texttt{yt}~\cite{Turk:2011}, \texttt{pynbody}~\cite{ascl:1305.002}, \texttt{numpy}~\cite{vanderWalt:2011}, \texttt{scipy}~\cite{Virtanen:2020},  \texttt{matplotlib}~\cite{Hunter:2007}, \texttt{ipython}~\cite{Perez:2007}, \texttt{astropy}~\cite{astropy:2013, astropy:2018}, \texttt{emcee}~\cite{ForemanMackay:2013} , \texttt{George}~\cite{Ambikasaran:2014}, \texttt{corner}~\cite{ForemanMackay:2016}

\appendix

\section{Thermal evolution in Nyx and Gadget simulations}
\label{sec:sim-thermal-cmp}

\begin{figure}
    \centering
    \includegraphics[width=\columnwidth]{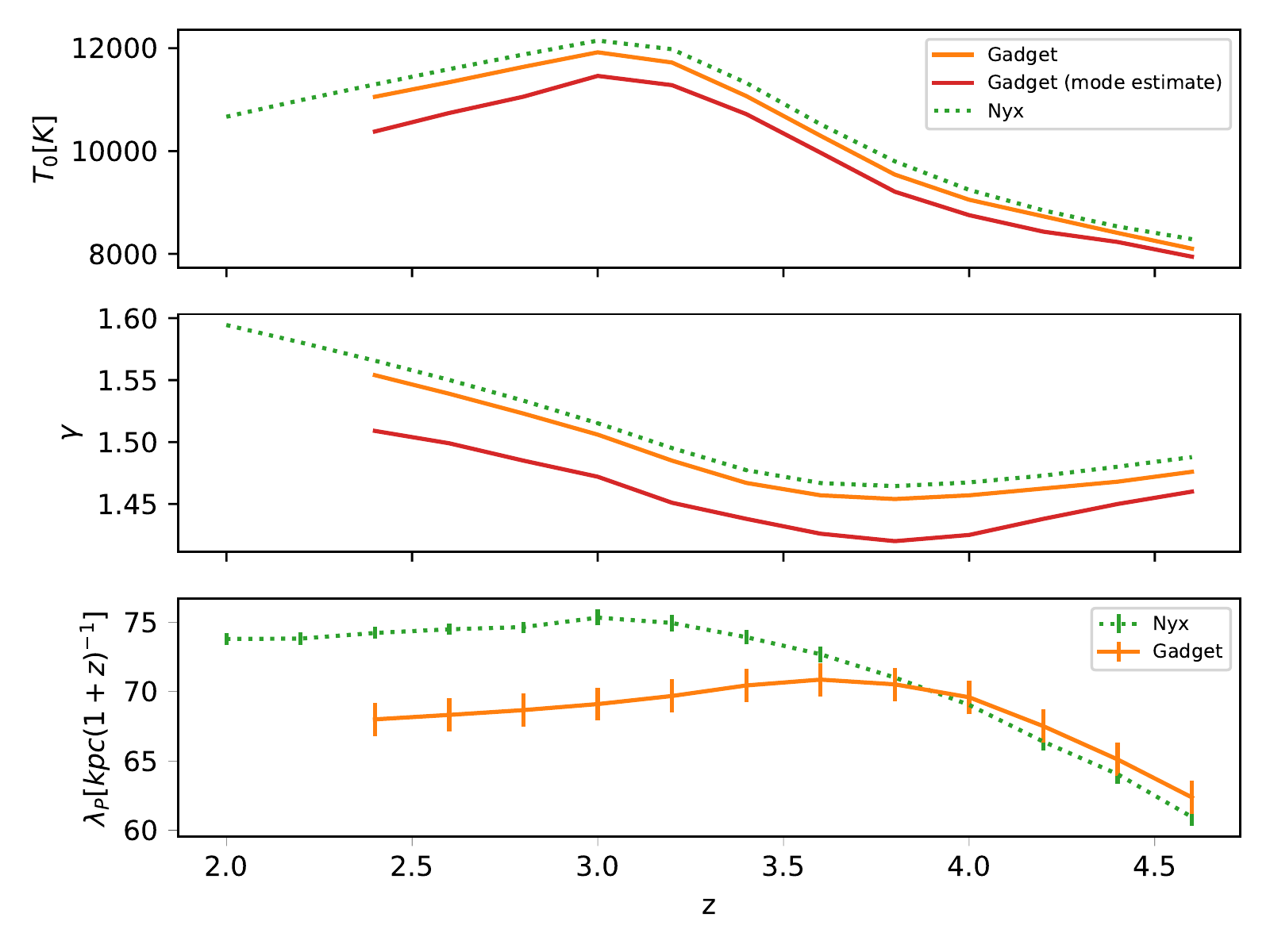}
    \caption{Thermal evolution quantified by a power law temperature density relation $T=T_0 \Delta^{\gamma-1}$ and a pressure smoothing scale $\lambda_P$ obtained from the different simulation codes using the same homogeneous UV background. For Gadget a second curve is shown based on first estimating the mode inside bins and then fitting the relation to the mode values (red curve).
    Note that while late time constraints on the thermal state are consistent between codes, there is a relatively large methodology dependence in the obtained parameters.
    For the Jeans scale in addition the analysis is relatively complex and depend on the exact fitting procedure (for Gadget the error bars are showing the scatter between 2 fitting procedures)}
    \label{fig:thermal-evo}
\end{figure}

A comparison of the thermal evolution in the fiducial Gadget and Nyx simulation runs is shown in \autoref{fig:thermal-evo}. The $T_0$ and $\gamma$ parameters are computed by adjusting the relation $T = T_0 \Delta^{\gamma-1}$ to the actual distribution of temperature versus baryon density in simulation boxes. We implement a volume-weighted fit of this relation in Nyx with a linear regression on the sample of all gas cells in $\log \Delta$ and $\log T$ with $0.1<\rho/\bar{\rho}<10$ and $T<10^5$~K (see e.g.~\cite{Lukic:2015}). For the Gadget simulation, we perform the equivalent fit, by weighting each particle with its inverse density (orange curves in Fig.~\ref{fig:thermal-evo}). We find that the Nyx and Gadget-derived values of $T_0$ and $\gamma$ are almost identical.

On a side note, we remind that in several references such as, for example, \cite{Borde:2013, Chabanier:2019}, $(T_0,\gamma)$ is estimated in a different way: modes of the $T-\rho$ distribution in bins of $\Delta$ are computed in the range $0.1<\rho/\bar{\rho}<1$, and a linear fit through these mode values is performed. The red curves in Fig.~\ref{fig:thermal-evo} illustrates the resulting thermal parameters by applying this procedure to our Gadget simulation. We find that the difference between both fitting methods is significant: at low redshift it can be as large as $\Delta T_0 \sim 600$~K and $\Delta \gamma \sim 0.05$. We have also tested a mass-weighted $(T_0,\gamma)$ fit for Gadget, and found that the $T_0$ and $\gamma$ estimates agree well with volume-weighted values.

The Jeans smoothing scale $\lambda_P$ is determined from simulation outputs, by fitting the cutoff in the power spectrum of the real-space \lya{} flux $F_{\rm real}$ as described in~\cite{Kulkarni:2015}. Here, $F_{\rm real}$ is the flux at each position in the simulation box (given its temperature and density), but neglecting redshift space effects, i.e. thermal broadening and peculiar velocities. The comparison between Nyx and Gadget (Fig.~\ref{fig:thermal-evo}, bottom) shows a similar evolution of $\lambda_P$ with redshift. For $z>3.5$, the values of $\lambda_P$ are fully compatible given the uncertainties associated to the fitting procedures. However the value of $\lambda_P$ during and after \HeII{} reionization is slightly larger for Nyx ($\lambda_P \simeq 74$~kpc at $z\sim 2.5$) than for Gadget ($\lambda_P \simeq 68$~kpc at the same redshift). This is consistent with the fact that the IGM temperature $T_0$ is slightly larger for Nyx over all the cosmic expansion.

\section{Matter distribution in Nyx and Gadget simulations}
\label{sec:sim-matter-dist-cmp}

\begin{figure}
        \centering
        \includegraphics[width=0.45\textwidth]{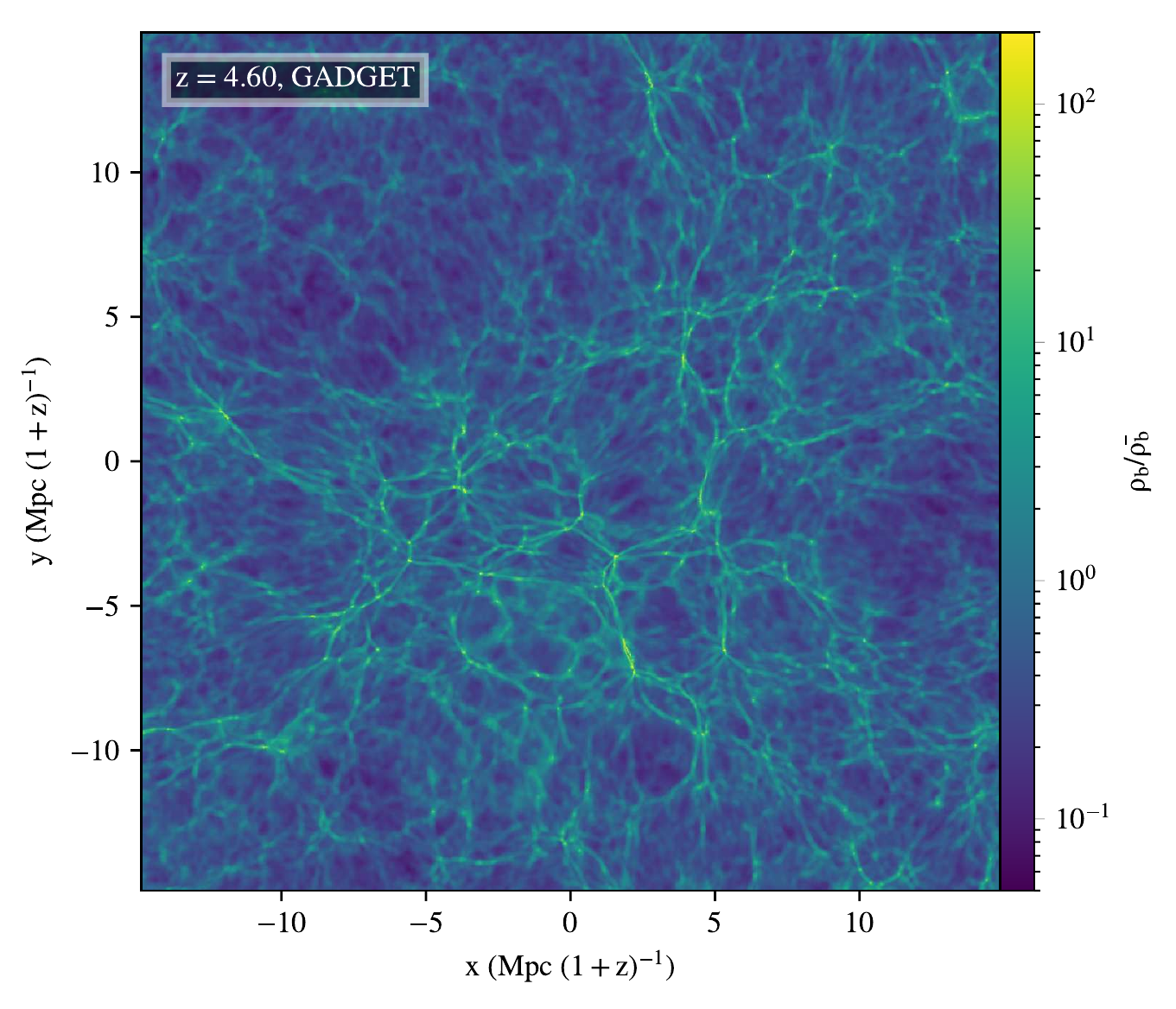}
        \includegraphics[width=0.45\textwidth]{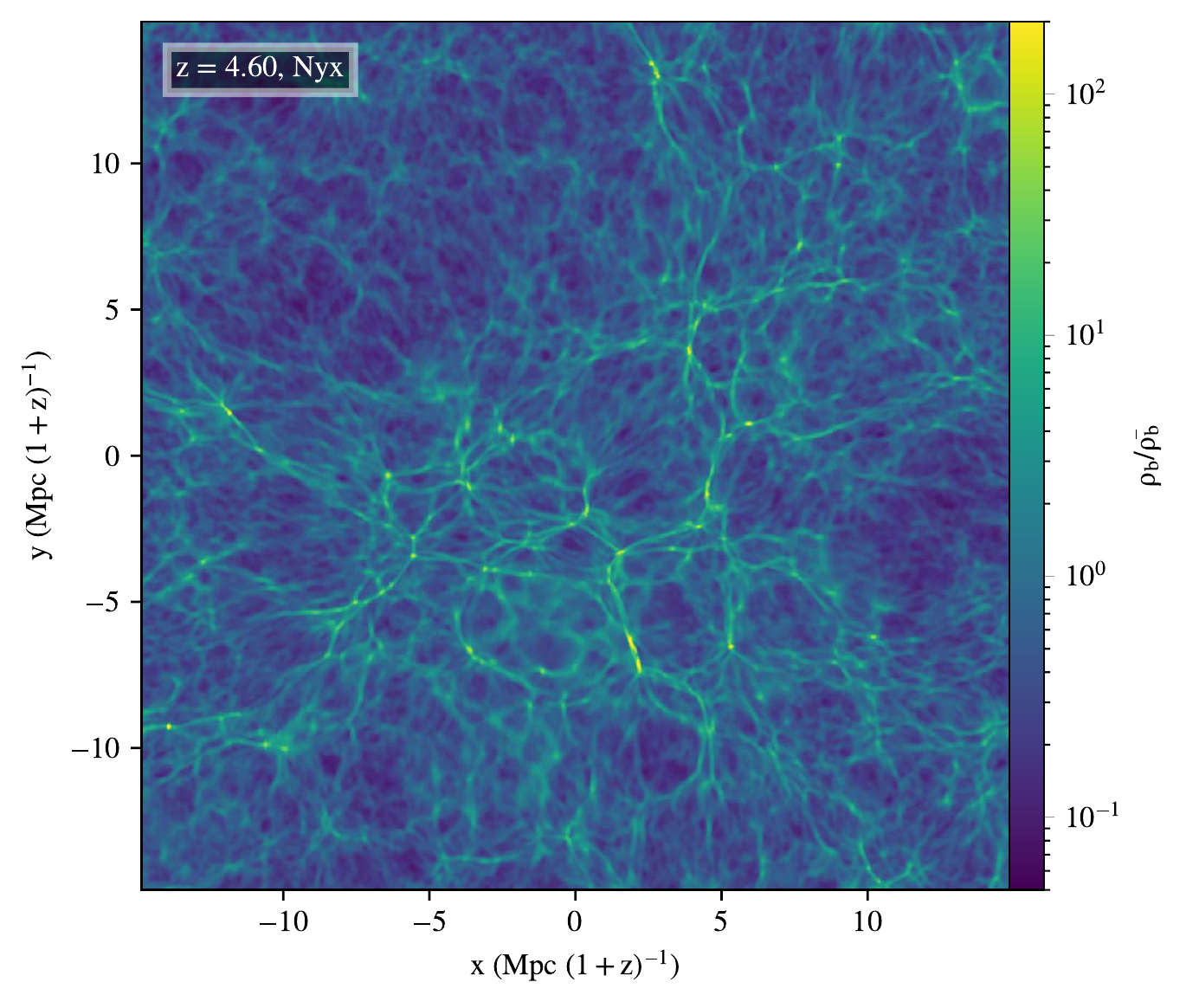}\\
        \includegraphics[width=0.45\textwidth]{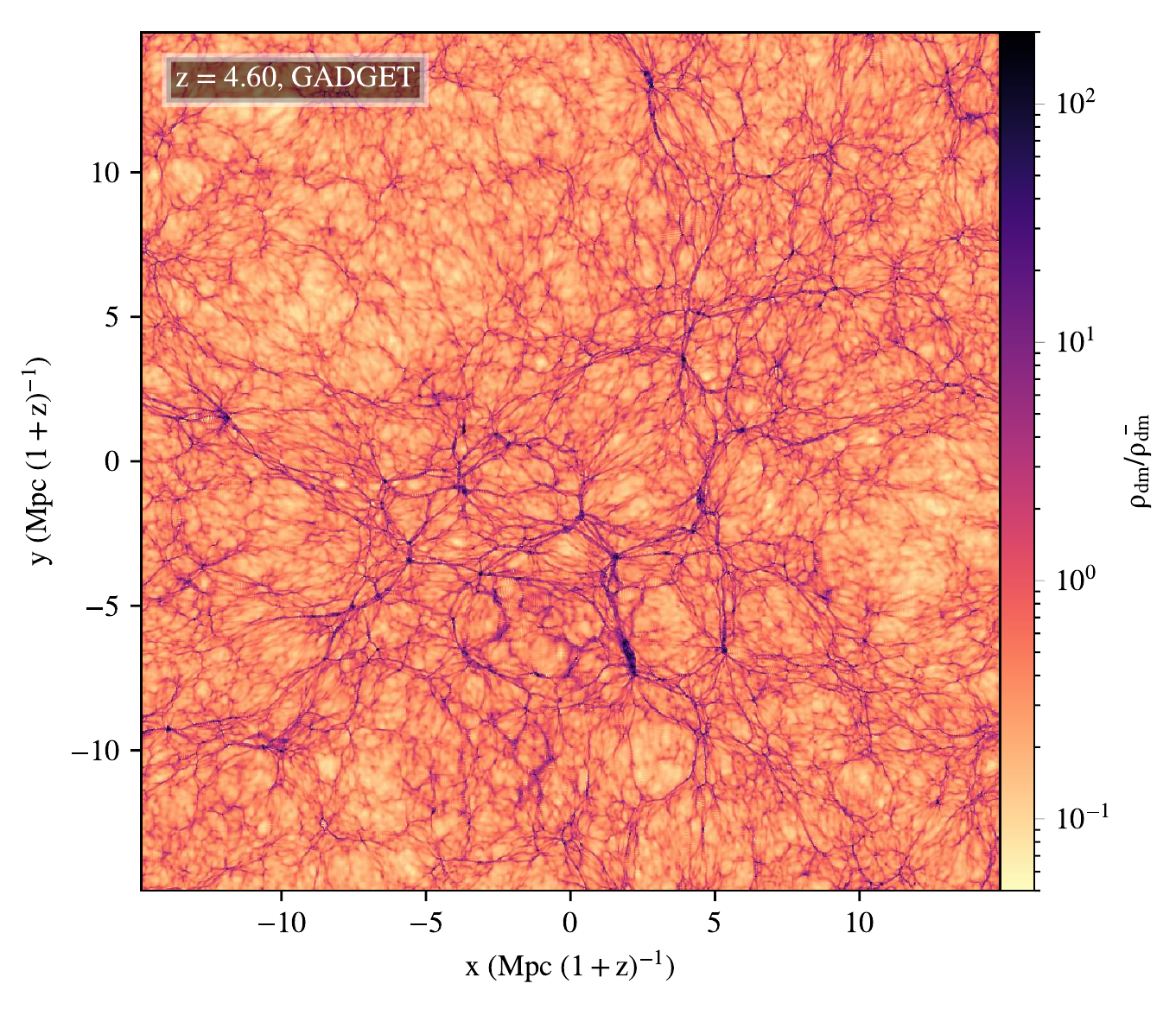}
        \includegraphics[width=0.45\textwidth]{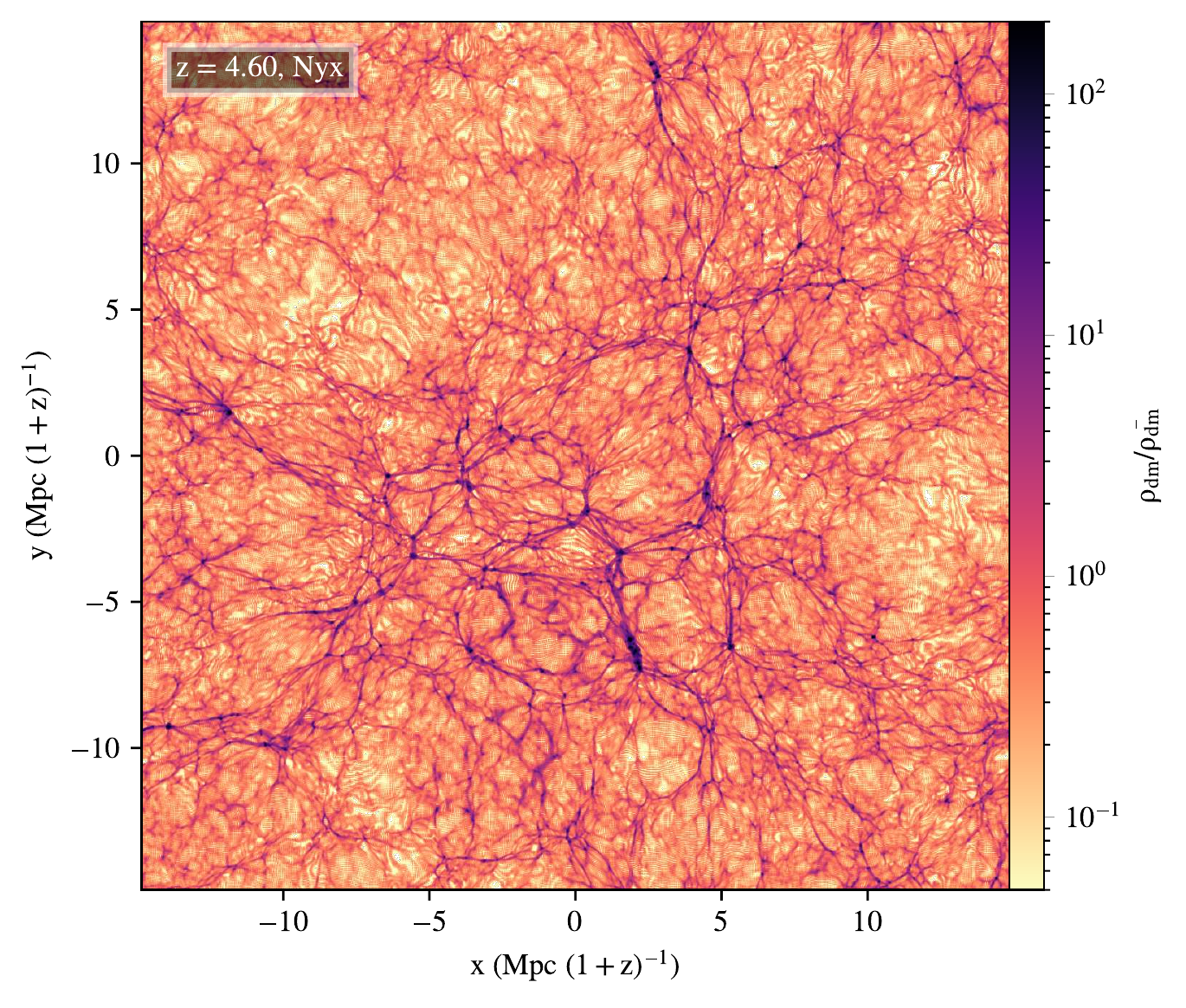}
        \caption{Slices at $z=4.6$ through the fiducial simulation outputs based on Gadget (left) and Nyx (right), for the densities of baryonic gas (top) and dark matter (bottom). }
        \label{fig:slice-density}
\end{figure}

We show a visual comparison of slices through the centers of the fiducial simulation boxes in \autoref{fig:slice-density}. This comparison clearly shows that the large-scale structure generated in both simulations are very similar, both for dark matter and the baryonic gas, as would be expected due to the very similar initial conditions.

\begin{figure}
        \centering
        \includegraphics[width=\textwidth]{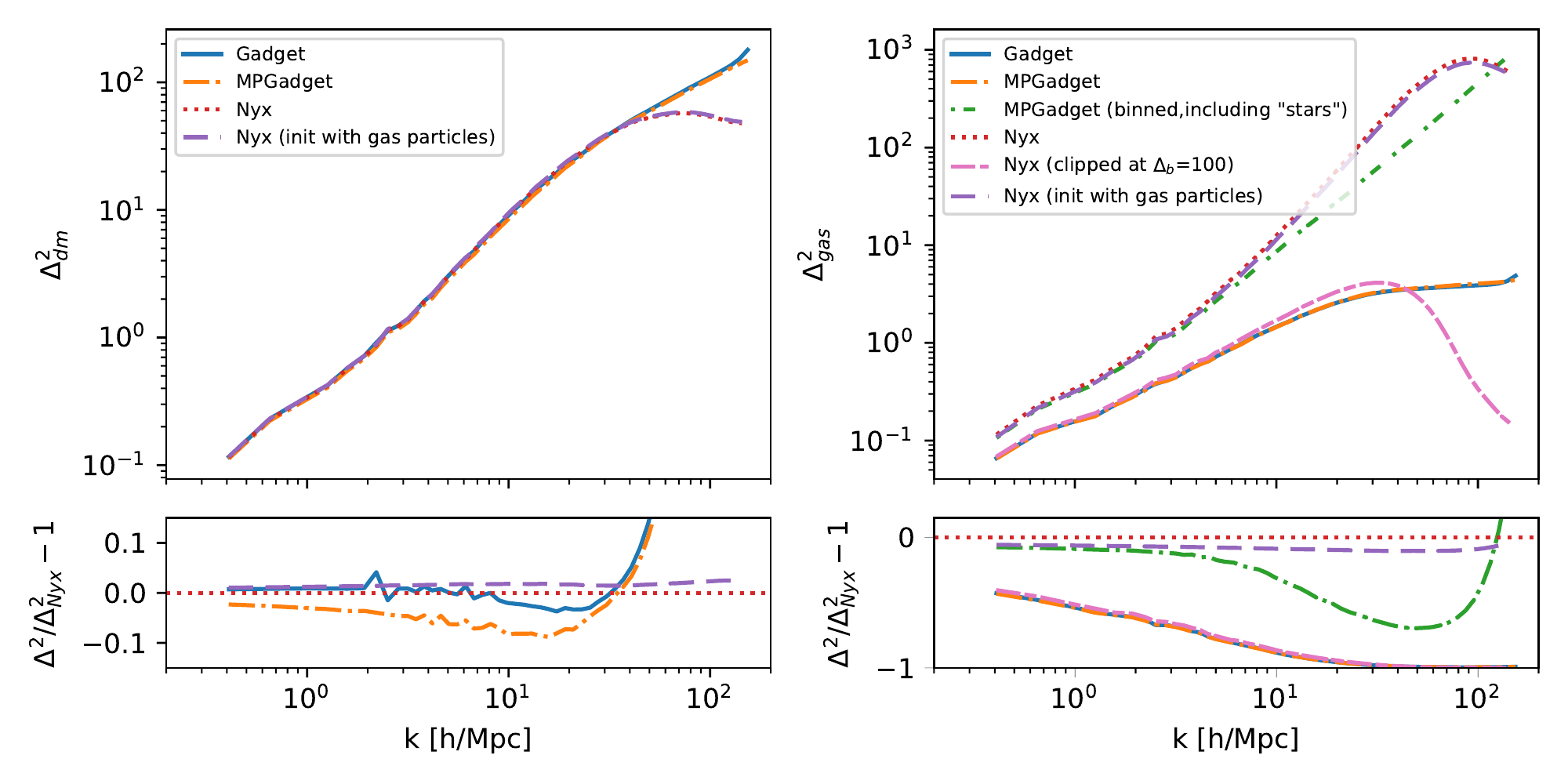}
        \caption{Power spectrum of dark matter (left) and baryons (right), from our Nyx simulation (red) compared to Gadget (blue) and MP-Gadget (orange) runs, at $z=3.8$. For MP-Gadget runs we show the baryon power both including (green) and excluding (orange) gas from high-density regions that was relabeled as stars in the QUICKLYA procedure. 
        Correspondingly, for Nyx the Baryon field is shown both including (red) and excluding (pink) the highest-density gas as well. 
        We also show Nyx results obtained by initializing the baryon grid from SPH particles (purple, see text).
        }
        \label{fig:matter power}
\end{figure}

We compare the matter power spectrum obtained from all 3 simulation codes in \autoref{fig:matter power}, for $z=3.8$. While there is some degree of redshift dependence within the results (e.g. an increasing impact of the small scales dark matter power differences, and decreasing impact of initial condition details with redshift), qualitatively the differences discussed in this section stay the same.
For dark matter clearly all 3 codes produce comparable results on large scales ($k<30\mathrm{h/Mpc}$) with the MP-Gadget run producing $\sim 5\%$ lower power than the other two. On smaller scales the Nyx run has significantly reduced power, which is due to the fixed grid size (no AMR was used).
Importantly, scales influenced by this are far smaller than scales accessible even with the high-resolution \lya{} forest observations.

For the baryons on the other side, the differences between SPH and Nyx code seem large on a first glance. However, most of the differences can be attributed to the QUICKLYA recipe which converts high density gas regions into star particles. Two approaches are thus used to provide a fair comparison between simulations:
\begin{itemize}
    \item We can calculate the full baryonic power spectrum from Gadget outputs, taking both gas and star particles into account (Fig.~\ref{fig:matter power}, cyan curve). We then recover a $\sim 5$~\% level agreement between both approaches for all large scale modes $k \lesssim 10 \,\hinvMpc$.
    \item On the other hand, the violet curve in Fig.~\ref{fig:matter power} shows the modified Nyx baryon power spectrum after the gas density was artificially clipped at $\Delta=100$ .
\end{itemize}
We can therefore conclude that the power spectra of gas density fluctuations agree to $\sim 5$~\% between different simulations, in the density range relevant to IGM and \lya{} studies.

Finally, as explained in Sect.~\ref{sec:sims}, while for the default Nyx run we generated both dark matter and baryons from the same transfer function, we also run an additional Nyx simulation initialized from the SPH (gas) particles. The resulting gas power spectrum is shown in Fig.~\ref{fig:matter power} (violet curve): the overall change is minor but improves the large scale agreement between Nyx and MP-Gadget baryons even more to the $2\%$-level. 

\section{Impact of the high-density component of the gas on the \lya{} flux power}
\label{sec:density-dependence}
\begin{figure}
    \centering
    \includegraphics[width=0.49\textwidth]{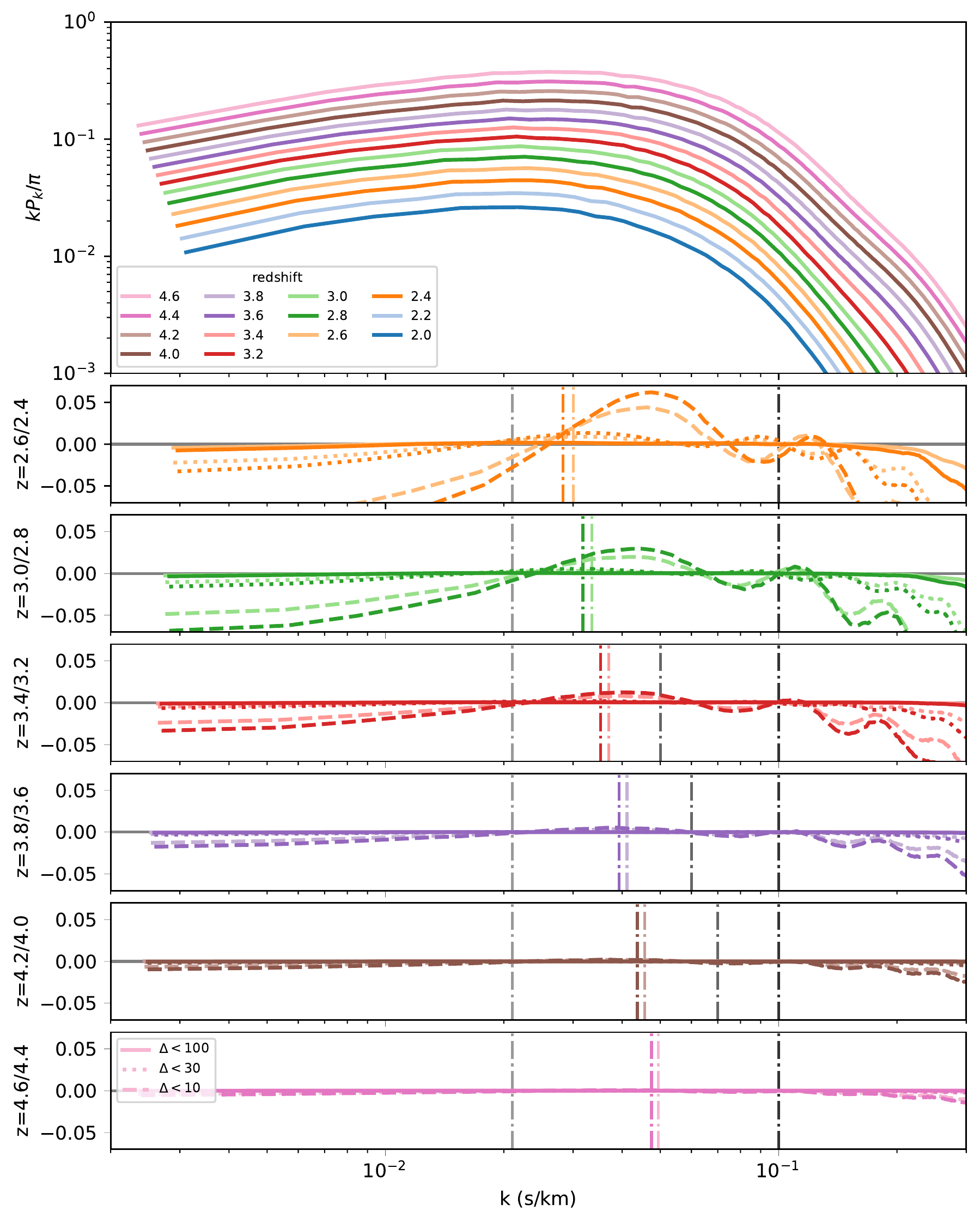}
    \includegraphics[width=0.49\textwidth]{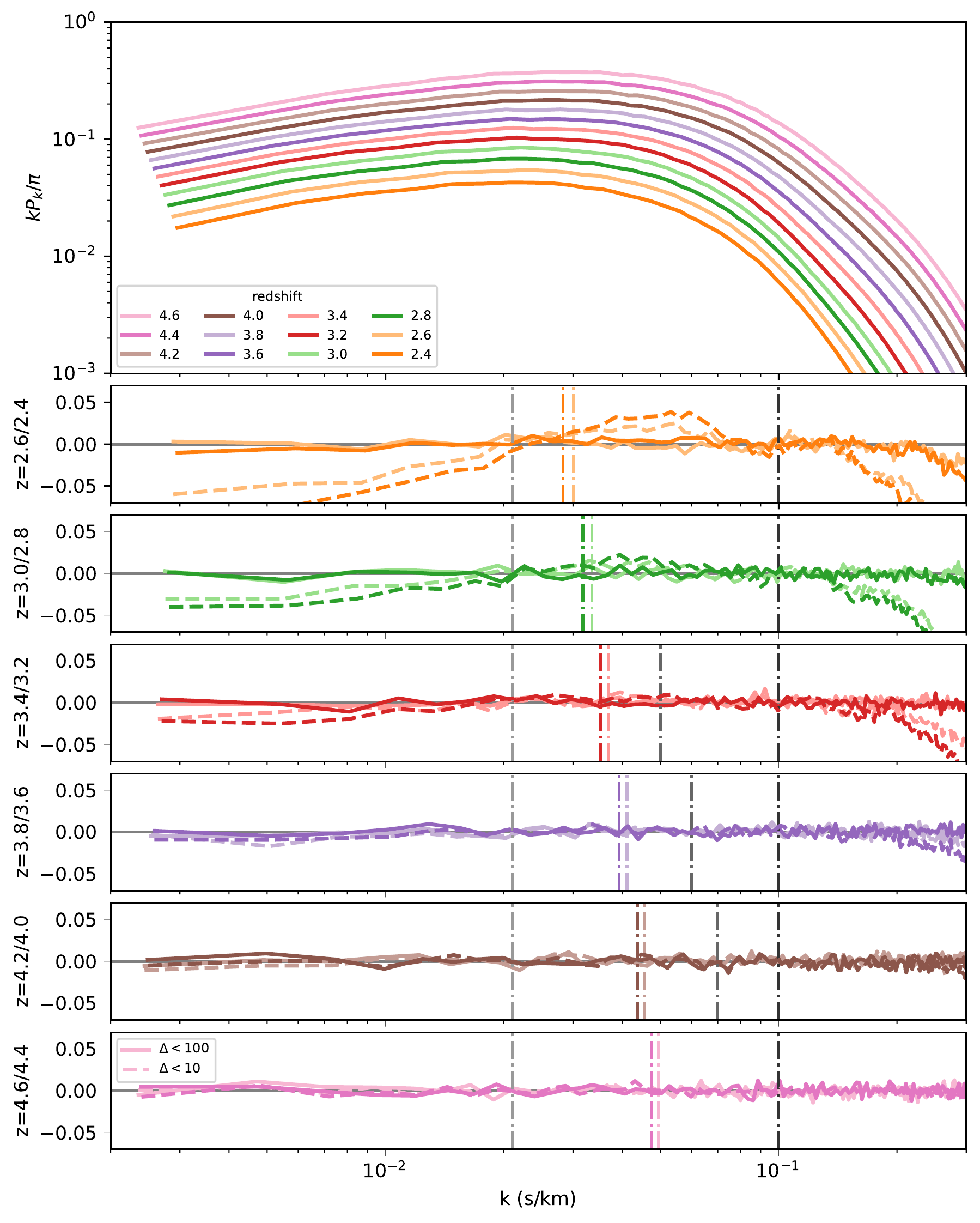}
    \caption{The same as \autoref{fig:PF_Gadget_vs_Nyx}, but now thresholding Nyx (left) and Gadget (right) runs at various gas overdensities ($\Delta_\mathrm{t}=10$ to 100 depending on linestyles), and showing differences to the non-truncated calculation.}
    \label{fig:Nyx_threshold_100}
\end{figure}

To estimate the relative contribution of gas at different overdensities to the flux power spectrum in the different codes, we recomputed the \lya{} spectra, but now truncating gas overdensities of $\Delta>\Delta_\mathrm{t}$ to the threshold density $\Delta_\mathrm{t}$, while keeping all other physical parameters identical. The results obtained when performing this truncation are shown in \autoref{fig:Nyx_threshold_100}. We first observe that the high-density gas has both small and large scale effects, and those are similar for both Gadget-3 and Nyx simulations. At high redshift ($z>3.6$), the flux power is almost not impacted by high gas density regions: even overdensities as low as $\Delta=10$ contribute to a sub-\% fraction of the power. At lower redshifts, $z\lesssim 3$, we can see that overdensities $\Delta\sim 30$ provide an up to $\sim 4\%$ contribution on large scales, whereas gas of $\Delta>100$ at no point creates a $>1\%$ change of power given the redshifts and scales observed via ground based \lya{} forest observations. 

For our purposes regarding the flux power we therefore conclude that it is never necessary to resolve halos of $\Delta>100$. We note that the conclusion would be strongly different concerning the baryon matter power, which is strongly affected by such density cuts. The reason for the low impact of high-density gas on the \lya{} flux power spectrum fundamentally lies in the saturation of \lya{} forest lines. Let us also remark that regions strongly affected by high-density gas are anyway typically excluded in \lya{} analyses, for several reasons like lacking resolution, accuracy of S/N corrections, and increase of metal absorption contamination (see e.g.~\cite{Walther:2018}).

\bibliographystyle{jcap}
\bibliography{citations_all}

\end{document}